\documentclass[12pt]{iopart}
\usepackage[T1]{fontenc} 
\usepackage{graphicx}
\usepackage{xcolor}
\usepackage{siunitx} 
\usepackage{multirow} 
\usepackage{bm}
\usepackage{pifont}

\expandafter\let\csname equation*\endcsname\relax
\expandafter\let\csname endequation*\endcsname\relax
\usepackage{amsmath}
\usepackage{amsfonts}

\newcommand{\exciting}{{\usefont{T1}{lmtt}{b}{n}exciting}}
\newcommand{\appropto}{\mathrel{\vcenter{\offinterlineskip\halign{\hfil$##$\cr\propto\cr\noalign{\kern2pt}\sim\cr\noalign{\kern-2pt}}}}}

\newcommand{\T}{\rule{0pt}{3.2ex}}
\newcommand{\B}{\rule[-1.8ex]{0pt}{0pt}}
\newcommand{\TTT}{\rule{0pt}{4.8ex}}
\newcommand{\BBB}{\rule[-2.4ex]{0pt}{0pt}}

\newcommand{\eg}{{\it e.g.}, }
\newcommand{\ie}{{\it i.e.}, }

\newcommand*\rot{\rotatebox{90}}

\begin{document}

\title[]{Advancing descriptor search in materials science: feature engineering and selection strategies}

\author{Benedikt Hoock$^{1,2}$, Santiago Rigamonti$^{1,*}$, and Claudia Draxl$^{1,2}$}
\address{$^1$ Institut f\"ur Physik and IRIS Adlershof, Humboldt-Universit\"at zu Berlin, Berlin, Germany}
\address{$^2$ Fritz-Haber-Institut der Max-Planck-Gesellschaft, Berlin, Germany}
\ead{srigamonti@physik.hu-berlin.de}

\begin{abstract}
A main goal of data-driven materials research is to find optimal low-dimensional descriptors, allowing us to predict a physical property, and to interpret them in a human-understandable way. In this work, we advance methods to identify descriptors out of a large pool of candidate features by means of compressed sensing. To this extent, we develop schemes for engineering appropriate candidate features that are based on simple basic properties of building blocks that constitute the materials and that are able to represent a multi-component system by scalar numbers. Cross-validation based feature-selection methods are developed for identifying the most relevant features, thereby focusing on high generalizability. We apply our approaches to an \textit{ab initio} dataset of ternary group-IV compounds to obtain a set of descriptors for predicting lattice constants and energies of mixing. In particular, we introduce simple complexity measures in terms of involved algebraic operations as well as the amount of utilized basic properties.
\end{abstract}

\maketitle

\section{Introduction} \label{sec:introduction}

On the steady search for advanced materials with tailored properties and novel functions, high-throughput screening has become a popular branch of materials research. Focusing on the computational side, the amount of materials data produced on workstations, compute clusters, and supercomputers is growing exponentially. This situation can be considered as a {\it big-data problem} as well as a chance -- the chance to learn from these data and obtain unprecedented insight, opening new routes for basic materials science and engineering. In this sense, \emph{big data of materials science} is all about the question: How to exploit the wealth of information, inherently inside the materials data? For successfully exploring the chemical compound space, new concepts need to be developed that allow us for extracting knowledge from the materials data, identifying trends, anomalies, and mechanisms.

This is where machine learning (ML), and, more general, artificial intelligence (AI) come into play. Learning from materials data in an automatized manner, as ML / AI may suggest, is, however, far from being trivial, since learning the behavior of the entire quantum-mechanical many-body problem is a too difficult task. To make reliable predictions, the most critical step is to find suitable descriptors, \ie physically meaningful parameters that describe a material's property best. This will enable us to develop and/or use AI tools to not only find optimal models but also to gain new insights into the underlying physical laws by relating properties and descriptors.

An important aspect of AI is to learn physical quantities that are computationally expensive and time-consuming from data that are easy to obtain. Predicting properties of bulk materials from atomic data or interface properties from bulk results are just two examples out of many possibilities \cite{Mueller2016}. For finite systems, atomization energies of relaxed molecules of similar kind have been successfully predicted within chemical accuracy \cite{Rupp2012}. Conversely, due to the long-range interactions, materials properties are much harder to predict. This has challenged both (i) the \emph{feature engineering}, \ie the design of convenient numerical representations most relevant to a property, and (ii) the approaches for feature selection, \ie the identification of the relevant features out of a possibly (very) large pool. Important examples for (i) include the generation of features for modeling the potential-energy surface of solids \cite{Behler2011, Bartok2013,Seko20141} or learning the cohesive energies of compounds for arbitary supercells \cite{Seko2017}.  More recently, crystal-graph representations applicable to arbitrary crystal structures, so called $n$-grams, were successfully employed, in combination with kernel ridge regression, to predict formation energies and band gaps of transparent conducting oxides \cite{Sutton2019}. An important example \cite{Ghiringhelli2015,Ghiringhelli2017}, advancing both (i) and (ii), concerns a {\it classical} classification problem, namely separating the binary octet semiconductors into whether they crystallize in rocksalt or zincblende structure. This problem has a long history in materials science (for an overview, see \cite{Butcher2013}). It has been solved by symbolic regressions via compressed sensing, in particular the LASSO+$ \ell_0 $ \cite{Ghiringhelli2015,Ghiringhelli2017} and SISSO \cite{Ouyang2018} approaches, and also by subgroup discovery techniques \cite{Goldsmith2017}. Other examples are crystal-structure classification problems using neural networks \cite{Ziletti2018}, the generation of structure-energy-property landscapes for molecular crystals \cite{Musil2018}, the prediction of vibrational free energies and entropies \cite{LeGrain2017}, the determination of the electrical breakdown field of insulators and the learning of the dielectric constant and band gap of polymers \cite{Ramprasad2017}, the prediction of band gap energies or glass-forming ability of crystalline and amorphous materials \cite{Ward2016}, and unified frameworks for predicting atomic-scale properties of molecules and solids \cite{Bartok20170}. 

In this work, we advance some of the above methodology to identify suitable descriptors in two main aspects. First, we show how meaningful, yet simple, candidate features can be constructed from basic properties of the materials that constitute the building blocks of multi-component systems. Second, we devise and assess strategies for model selection. Thereby we also consider the predictive power of the model by making use of cross-validation for both model selection and validation. We demonstrate our approaches with the example of group-IV zincblende ternary alloys, for which we aim at predicting lattice constants and energies of mixing. We use an \textit{ab initio} dataset obtained by density-functional theory (DFT) on the level of the local-density approximation (LDA) which yields reasonable accuracy for the quantities of interest in our work, \ie $\sim1$\% for lattice constants and $\sim4$\% for energies of formation \cite{Yin1981}. Since the properties of such alloys are sensitive to both the composition and the atomic arrangement, this poses a challenge to the construction of appropriate candidate features. Generating a feature space for disordered structures has, so far, not been tackled systematically in the context of symbolic regression.  Focusing on the two learning tasks, we first determine a set of basic physical properties and mathematical operations for generating the feature space that provides a starting point for applying compressed-sensing methodology  \cite{Ghiringhelli2015,Ghiringhelli2017}. In particular, we examine the quality of the models in terms of a measure of descriptor complexity, where the included primary features as well as the descriptors' algebraic form are considered. Further, we compare the different model-selection strategies based on cross validation as well as LASSO+$\ell_0$ and SISSO, that we use for identifying the most meaningful descriptors. Finally, we present a set of \textit{optimal} descriptors as a solution to the two learning tasks and discuss their interpretability.

\section{Background} 
\label{sec:background}

\subsection{Building a linear model} \label{subsec:building_a_linear_model}
As very often in a learning task, one wants to find a linear model that allows one to predict a material property $P$ based on some input data $\mathbf{D}$. More specifically, the aim is to obtain, in the context of compressed sensing, a small subset of input vectors from the potentially huge matrix $\mathbf{D}$ which both gives an accurate prediction $\hat{P}$ of the property and desirably can be interpreted in a physically meaningful way. This subset is called $\Omega$-dimensional descriptor $\mathcal{D}$, where $\Omega$ is the number of considered input vectors. Formally, the linear model can be written as
\begin{equation}
\hat{P}_s=\mathbf{D}_s\cdot \mathbf{c}=\sum_{i=1}^\Omega F_{sk_i}c_{k_i}.
\label{eq:model}
\end{equation} 
The descriptor ${\cal D}$ can then be defined as a set of features ${\cal F}_k$ which map a material $s$ to the corresponding set of real numbers, \ie  
\begin{eqnarray}
{\cal F}_k: s \rightarrow F_{sk} \in {\rm I\!R},  \\
{\cal D}: s \rightarrow \mathbf D_{s}=(F_{sk_1},...,F_{sk_\Omega}) \in {\rm I\!R}^\Omega.\label{eq:dmat}
\end{eqnarray} 
The components of the coefficient vector $\mathbf{c}=(c_{k_1}, c_{k_2}, ..., c_{k_\Omega})$ are found through fitting procedure to a training set $S=\left\lbrace (\mathbf{D}_1,P_1),...,(\mathbf{D}_N,P_ N)\right\rbrace$  where $N$ is the number of training samples. The property $P_s$ in each data-point $(\mathbf{D}_s,P_s)$ must be known beforehand, either from calculations or measurements. The task of finding the appropriate descriptor is achieved by first constructing a potentially huge pool of $M$ candidate features $\left\lbrace {\cal F}_1,..., {\cal F}_M \right\rbrace$ and then learning from this pool the $\Omega$ features yielding the \emph{best} model.

\subsection{Compressed sensing methods} 
\label{subsec:compressed_sensing_methods}

In order to extract only a few highly relevant features from the plenty of candidates contained in $\mathbf{D}$ (see Section \ref{sec:construction_of_features}), we make use of two different feature-selection methods $\mathcal{M}$, namely LASSO+$\ell_0$ \cite{Ghiringhelli2015,Ghiringhelli2017} and SISSO \cite{Ouyang2018} that are based on compressed-sensing theory and are briefly sketched in the following. Utilizing these two methods, we have developed various model-selection strategies, based on cross-validation, which aim at generalizability of the resulting models. They will be described in Section \ref{sec:model_selection_strategies}.

\subsubsection{The LASSO+$\ell_0$ approach} 
\label{subsubsec:LASSO}

Mathematically, in compressed sensing, the search for a descriptor is expressed by the optimization problem
\begin{equation}
\underset{\mathbf{c}}{\textrm{argmin}} \left\Vert \mathbf{P} -  \mathbf{D} \mathbf{c} \right\Vert_2^2 + \lambda \left\Vert \mathbf{c} \right\Vert_0 .
\label{eq:ell0}
\end{equation}
Here, the input matrix $\mathbf{D}$ is of size $N\times M$ containing the elements $F_{sk}$. $\left\Vert \mathbf{v} \right\Vert_p$ denotes the $\ell_p$-norm of $\mathbf{v}$, defined as $\left(\sum_{i=1}^N |v_i|^p\right)^{1/p}$. 
For the special case $p=0$, $\left\Vert \mathbf{v} \right\Vert_0$ is defined as the number of non-zero elements of $\mathbf{v}$, \ie $\# \left\lbrace i \textrm{\, for which \,} v_i\neq 0\right\rbrace$. The solution of equation \ref{eq:ell0} yields the coefficients $\mathbf{c}$ that minimize the mean square error (MSE) between the target $\mathbf{P}$ and the prediction of the model $\mathbf{D}\mathbf{c}$, on the condition that the vector $\mathbf{c}$ is \emph{sparse}, \ie $ c_i \ne 0 $ for a few $ i \in \left\{1,\dots,M\right\} $ only. The parameter $\lambda$ is a positive real number. Higher values of $\lambda$ yield sparser models. If the solution has exactly $\Omega$ non-zero coefficients $c_i$, the associated subset of features $\lbrace {\cal F}_i \rbrace$ is an $\Omega$-dimensional descriptor.

A solution to this problem consists of extracting all possible descriptors ${\cal D}$ of dimension $\Omega$ from the pool of features and selecting the one yielding the lowest MSE. This is, however, practically impossible due to the combinatorial explosion of the number of possible descriptors at increasing $\Omega$ ($NP$-hard problem \cite{Arora2009}). To overcome this computational difficulty, in \cite{Ghiringhelli2015} and \cite{Ghiringhelli2017} an approximation to the $\ell_0$-problem has been proposed, consisting of two steps. First, one solves the convex optimization problem
\begin{equation}
\underset{\mathbf{c}}{\textrm{argmin}} \left\Vert \mathbf{P} -  \mathbf{D} \mathbf{c} \right\Vert_2^2 + \lambda \left\Vert \mathbf{c} \right\Vert_1.
\label{eq:ell1}
\end{equation}
Compared to equation \ref{eq:ell0}, this equation contains the $\ell_1$-norm (also called \emph{Manhattan} norm) of the coefficient vector $ \mathbf{c}$. This approach is known as the least absolute shrinkage and selection operator (LASSO) \cite{Tibshirani1996} and is the best convex proxy to equation \ref{eq:ell0}. It yields a subset of moderate size $\tilde{M}$, containing the most relevant features.  Second, equation \ref{eq:ell0} is then solved exactly by going through all combinations of features, in the subset obtained in the first step. As a result the best $\Omega $-dimensional descriptor out of $\tilde{M}$ are obtained.

\subsubsection{The SISSO approach.} \label{subsubsec:SISSO}
Although the LASSO+$\ell_0$ approach has been successfully applied, \eg to classification problems \cite{Ghiringhelli2015}, it turned out to experience difficulties. This happens on the one hand, at large feature matrix $\mathbf{D}$ where the procedure becomes computationally very costly, and, on the other hand, if $\mathbf{D}$ contains highly correlated features. As a solution to this undesirable situation, the {\it SISSO} method was proposed \cite{Ouyang2018} as an alternative approach to tackle the $\ell_0$-problem, also enabling to work with much larger feature spaces.

Similar to LASSO+$\ell_0$, SISSO combines two feature-selection methods. First, the \emph{sure independence screening} method (SIS) \cite{Fan2008} is applied to reduce the number of candidate features to a drastically smaller subset. Second, the \emph{sparsifying operator} (SO) further reduces the dimensionality, eventually determining the $\Omega$-dimensional descriptor. In more detail, SIS determines a first lower-dimensional subspace $\bm{S}_1$ of features by choosing those being mostly correlated with the target $\mathbf{P}$. Here, the \emph{Pearson correlation coefficient} is used as a measure. SO will then find an optimal 1D descriptor ${\cal D}_{\textrm{1}}$ from $\bm{S}_1$ (for 1D, this coincides with the first ranked by SIS). After this, SIS again is used to find a second subspace $\bm{S}_2$ that now contains the features of highest correlation with the residual $\mathbf{P} - \mathbf{D}^{(\textrm{1})} \mathbf{c}_{\textrm{1}} $ ($\mathbf{D}^{(\textrm{1})}$ being the feature matrix with rows given by Eq.(\ref{eq:dmat}) with ${\cal D}\rightarrow{\cal D}_{\textrm{1}}$). Like above, SO determines the best 2D descriptor -- but now from the union of subspaces $ \bm{S}_1 \cup \bm{S}_2 $. Thus, this procedure yields a sequence of almost orthogonal subsets and is iterated until a threshold in accuracy or in descriptor dimension $\Omega$ is reached. For the selection operator SO, several options are discussed in \cite{Ouyang2018} that may put restrictions on the feasible size of the subspaces. In this work, we choose SO($\ell_0$) to solve the $\ell_0$ problem exactly on the subset. Thus, the essential difference between LASSO+$\ell_0$ and SISSO is the set onto which $\ell_0$ is operated, being LASSO or SIS, respectively.

\section{Dataset of group-IV ternary alloys} 
\label{subsec:ternaries}

\begin{figure}[htb]
\centering 
\includegraphics[width=0.8\textwidth]{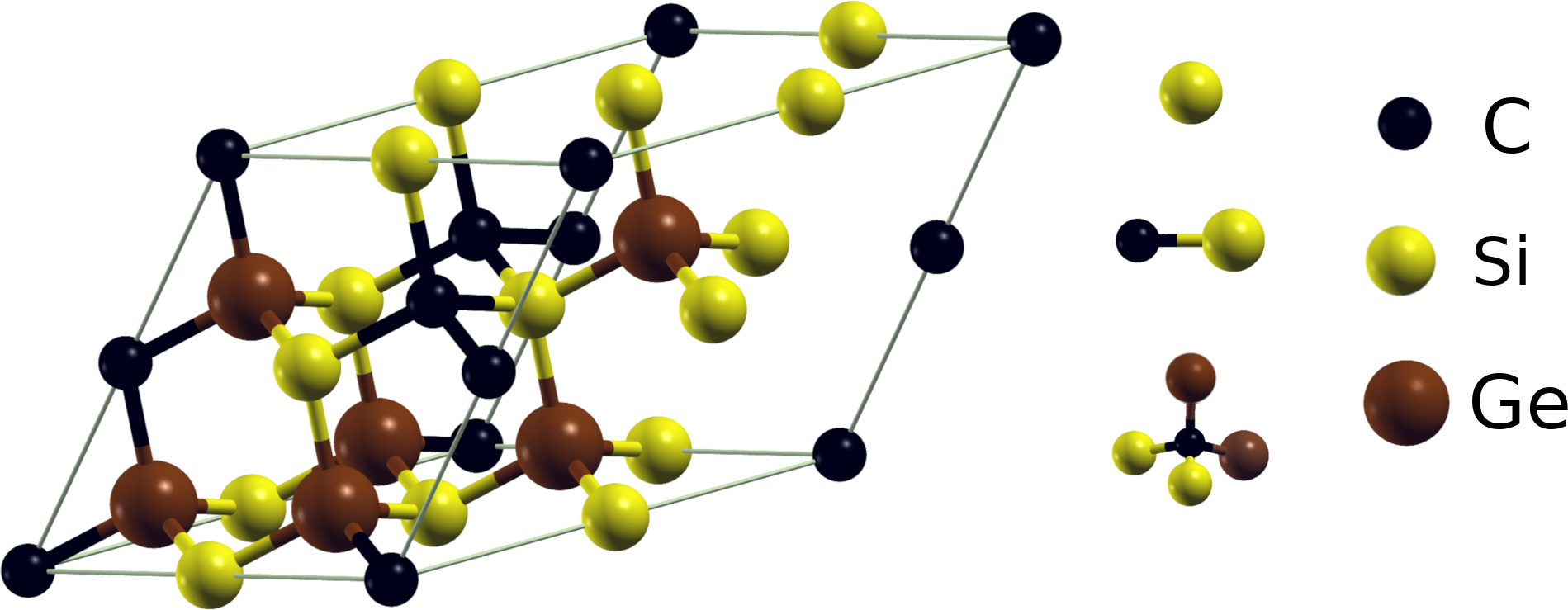}
\caption{Supercell representing a configuration of the ternary structure $\textrm{Si}_6\textrm{C}_4\textrm{Ge}_6$. Each atom is surrounded by four next neighbors in tetrahedral coordination. Right to it are examples of building blocks considered for the feature space: single atom (top), dimer (middle) and tetrahedron (bottom).}
\label{fig:ternary_and_building_blocks}
\end{figure}
We consider group-IV ternary compounds that are composed of three out of the four elements carbon, silicon, germanium, and tin. This material class is of great interest because of its potential electronic, photoelectronic, and photovoltaic applications \cite{Ventura2015,Fischer2015,Wendav2016}. Here, we focus on ternary structures that can be described by a 16-atom supercell (see figure \ref{fig:ternary_and_building_blocks}) as constructed by doubling the two-atomic zincblende parent lattice in each direction. The concentrations $x$, $y$, $1-x-y$ of types $A$, $B$ and $C$ are restricted to $0 < x,y \leq 0.5$ and $0 < 1 - x - y \leq 0.25 $. The code \emph{enumlib} \cite{Hart2008} was used to determine all symmetrically distinct structures that satisfy these restrictions. These leads to a total of 504 different configurations, out of which 445 were used \cite{anote}.

In general, a certain composition $(x,y)$ can be realized by several symmetrically inequivalent atomic arrangements, \ie $(x,y)$ does not define a structure $s$ uniquely. Typically, not only the composition but also the configuration has an impact on the physical properties. This needs to be reflected in the construction of the feature space. Specifically, for modeling the alloys' properties $P(x,y)$, this requires to go beyond pure linear interpolation schemes like Vegard's law \cite{Vegard1921}

\begin{equation}
P_\mathrm{lin}(x,y)=x P_A+y  P_B + (1-x-y)  P_C.
\label{eq:VegardsLaw}
\end{equation}
or interpolation of the properties of the binary compounds $P_{AB}$, $P_{AC}$, and $P_{BC}$ \cite{Adachi2009}. Instead, configuration-aware features must be introduced.

The dataset used in this work was calculated with the all-electron full-potential code \exciting\ \cite{Gulans2014} using the Perdew-Wang exchange-correlation potential \cite{PerdewWang1992}. Muffin-tin radii of 1.45, 1.65, 1.8, and 1.9 $a_0$ were employed for C, Si, Ge, and Sn, respectively, and the basis-set size was determined by the dimensionless parameter $R_{MT,{\textrm{min}}} \left| \mathbf{G} + \mathbf{k} \right|_{\textrm{max}} = 7.0$ for C, \ie the species with the smallest sphere. For the density and the potential, a plane-wave cut-off $\left| G \right|$ of $12.0 \, a_0^{-1}$ was applied. Brillouin-zone integrations were carried out on a $4 \times 4 \times 4$ $\mathbf{k}$-point mesh. The atomic positions were relaxed until all forces were converged better than $3.0 \cdot 10^{-4} \, \textrm{Ha} / a_0$. The lattice constants $a$ were determined by fitting the Murnaghan equation of state \cite{Murnaghan1944} to five values of the volume varying by up to $\pm 4 \%$ around an initial guess from Vegard's law. A final atomic relaxation of the equilibrium supercell yielded the groundstate energy $E(s)$ for each structure. The energy of mixing per atom, is obtained as $E_{\mathrm{mix}} = [E(s) - E_\mathrm{lin}(x,y)]/16$, where $E_\mathrm{lin}(x,y) = x \cdot E_A + y \cdot E_B + (1-x-y) E_C$ is the linearly interpolated groundstate energy according to equation \ref{eq:VegardsLaw}, and $E_A$ ($E_B$, $E_C$) is the energy of the pristine crystal in the diamond structure for species $A$ ($B$, $C$).

\begin{figure}[htb]
\includegraphics[width=0.9\textwidth]{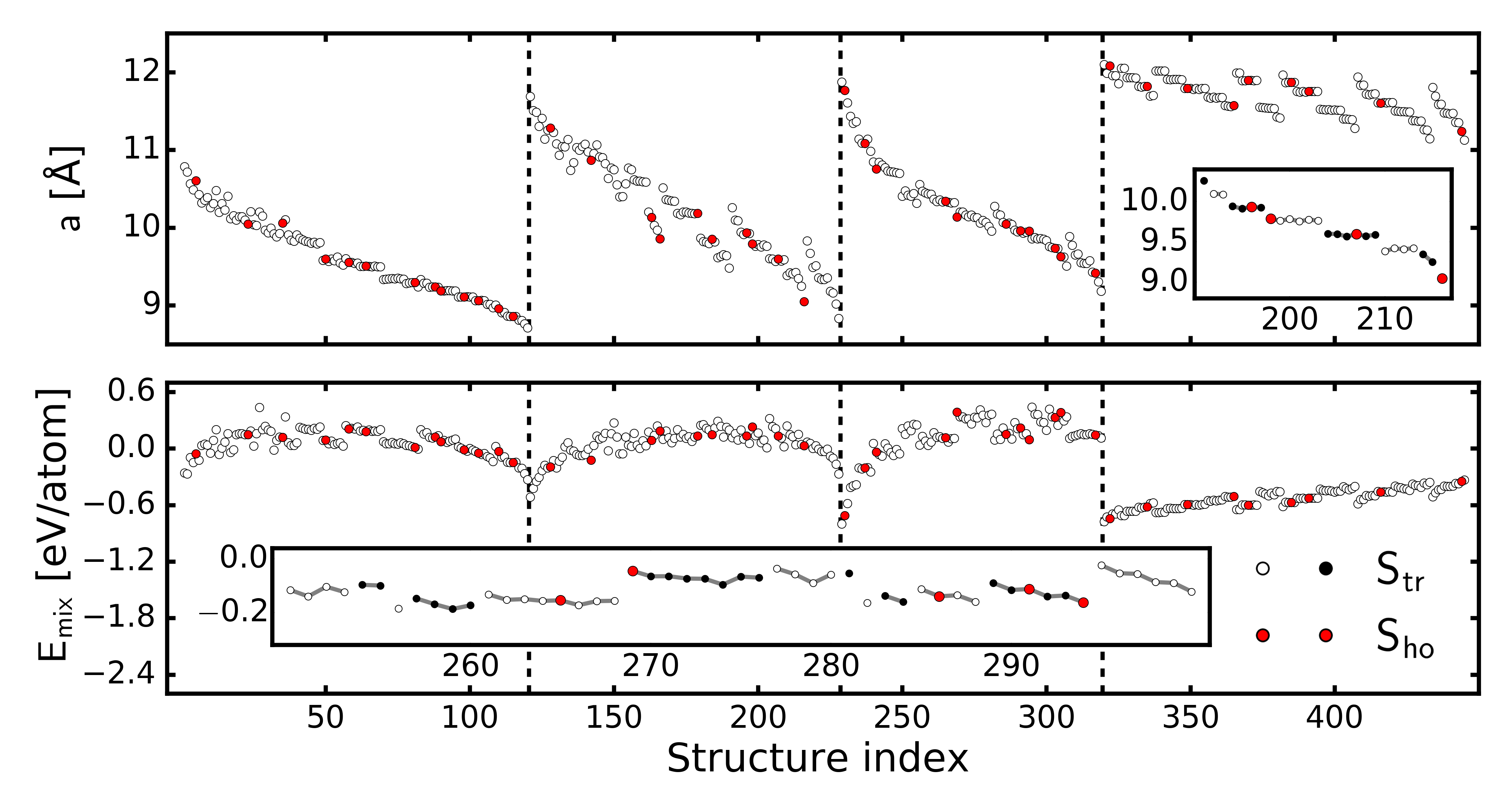}
\caption{Lattice constant, $a$, (top) and energy of mixing, $E_{\mathrm{mix}}$, (bottom) of the 445 structures in the dataset. The dotted vertical lines separate groups of structures containing the same three elements, \ie (C,Si,Ge), (C,Si,Sn), (C,Ge,Sn), and (Si,Ge,Sn) from left to right. Data points used for model selection, $S_{\textrm{tr}}$, are indicated by white and black dots, data points in the hold-out set $S_{\textrm{ho}}$ (different for $a$ and $E_\mathrm{mix}$) are marked in red. The insets illustrate the impact of the atomic arrangement. Here, groups of fixed composition are connected by a gray line.}
\label{fig:data}
\end{figure}

For these data, figure \ref{fig:data} shows the two target properties $a$ (top) and $E_\mathrm{mix}$ (bottom). The structures are sorted by ascending concentration of atom type $A$, then $B$, and then $C$ where $A$, $B$, and $C$ are sorted by ascending nuclear number. The insets, showing several groups of fixed composition, visualize the dependence on the arrangement as mentioned above. This dependence can be as large as 0.22\,\AA\ for the lattice constant and $\SI{0.41}{eV}$ for $E_{\mathrm{mix}}$, considering the entire data set. Further analyzing the dataset, we find that the relaxed atomic positions frequently deviate quite markedly from those of the ideal zincblende structure. For some structures, the bond lengths can differ by up to 10\%. We note that even bond lengths within the same bonding motif, for example bonds between silicon and germanium, and within the same structure may take different values. 

The data used in this work are available for download at the NOMAD Repository \cite{Draxl2019,Draxl2018}, DOI https://doi.org/10.17172/NOMAD/2022.05.20-1.

\section{Construction of feature spaces} 
\label{sec:construction_of_features}

How to build a feature space for materials-science problems, has been extensively described in \cite{Ghiringhelli2015,Ghiringhelli2017}. In short, one first defines a set of {\it primary features}, $\lbrace f_k \rbrace$, that are basic physical quantities of lower complexity than the target properties, \ie, requiring (much) lower computational cost than what is needed for computing the desired property. From this, one builds the full feature space by applying mathematical operators. In \cite{Ghiringhelli2015,Ghiringhelli2017}, binary semiconductors with up to only two atoms in the unit cell were considered, hence, a representation with site-dependent $f_k$'s was suitable. For the larger unit cells with disorder considered in this work, such representation is not convenient, since it \textit{i)} is not invariant under symmetry operations of the underlying crystal structure, and \textit{ii)} relies on a fixed unit (super) cell. Therefore, we extend the above described approach \cite{Ghiringhelli2015,Ghiringhelli2017} to make it applicable to supercells of arbitrary size with disorder. Essentially, we use the site-dependent $f_k$'s to define averaged primary features $F_k$. These are invariant under symmetry operations of the underlying lattice and also independent of the supercell size. From these one then builds the full feature space similar to \cite{Ghiringhelli2015,Ghiringhelli2017}.

Starting from the primary features, as displayed in tables \ref{tab:features_a} and \ref{tab:features_e}, these quantities are transformed into scalar primary features that characterize a sample material with a potentially complex structure as a whole. To obtain this, we perform an average of the $\lbrace f_k \rbrace$ over the crystal supercell as
\begin{equation}
F_{sk} = \frac{1}{m}\sum_{\gamma \in s} f_{k}(\gamma)
\label{eq:def_ave}
\end{equation}
where the sum runs over all building blocks $ \gamma $ of a certain type, and $ m $ is the number of these building blocks $ \gamma $ in the structure $s$. In the applications discussed in Section \ref{sec:application}, for instance, we employ averages over all single atoms, all pairs (or, equivalently, all bonds) and all tetrahedral clusters in the  16-atom supercell (figure \ref{fig:ternary_and_building_blocks}, right). As an example, the equilibrium dimer distance $ f_k = d_{2di} $ from the pool of primary features is summed over all pairs of nearest neighbors $\langle i,j \rangle$ in the supercell and divided by the total number of pairs in these structures ($m=32$). This averaging procedure can be viewed as a generalization of Vegard's law.

\begin{table}[h]
\caption{Primary features $f_k$ and different feature spaces $X_{a,i}$ ($i$=1-3) used for learning the lattice constant $a$. The crosses indicate that a quantity is included in $ X_{a,i} $. $at$, $es$, $bi$, $1di$, $2di$, $tet$ used for atomic, elemental solid, binary solid, 1-atomic dimer, 2-atomic dimer, and tetra, respectively. }\label{tab:features_a}
\begin{tabular}{ l | l | c| c| c}
  \hline\hline
Feature ($f_k$)	& Description & $ X_{a,1} $ & $ X_{a,2} $ & $ X_{a,3} $ \\
\hline
\multicolumn{5}{l}{Properties of mono-atomic systems}\\
\hline
$Z$ 	& Atomic number &  \ding{53}  &  \ding{53}    &  \ding{53}  \\
$P$   & Period (in periodic table) &  \ding{53}  &  \ding{53}    &  \ding{53}  \\
$r_s$ & $s$-orbital radius &  \ding{53}  &  \ding{53}  &     \ding{53}  \\
$r_p$ & $p$-orbital radius &  \ding{53}  &  \ding{53}  &     \ding{53}  \\
$r_d$ & $d$-orbital radius &  \ding{53}  &  \ding{53}  &     \ding{53}  \\
$a_{es}$ & Lattice constant of elemental solid &    &  \ding{53}     &  \ding{53}  \\
$d_{1di}$ & Bond length of mono-atomic dimer &    &  \ding{53}  &     \ding{53}  \\
\hline
\multicolumn{5}{l}{Properties of two-atomic systems}\\
\hline
$ a_{bi} $ & Lattice constant of binary material &    &  \ding{53}     &  \ding{53}  \\
$ d_{2di} $ & Bond length of bi-atomic dimer &    &  \ding{53}     &  \ding{53}  \\
\hline
\multicolumn{5}{l}{Properties of tetrahedra}\\
\hline
$ d_{tet} $ 	& Bond length &    &      &  \ding{53}  \\
\hline\hline
\end{tabular}
\end{table}

\begin{table}[h]
\centering
\caption{Primary features $f_k$ and different feature spaces $X_{E,i}$ ($i$=1-3) used for learning the energy of mixing, $E_\textrm{mix}$. The quantities $Z$, $P$, $r_s$, $d_{2di}$ and $d_{tet}$ are the same as in table \ref{tab:features_a}. In addition, the Kohn-Sham band-gap and the energy of formation are considered here.
\label{tab:features_e}}
	\begin{tabular}{ l | l | c| c| c}
		\hline\hline
		Feature	& Description & $ X_{E,1} $ & $ X_{E,2} $ & $ X_{E,3} $\\
		\hline
		\multicolumn{5}{l}{Properties of mono-atomic systems}\\
		\hline
		$ Z $ 	& Atomic number &  \ding{53}  &  \ding{53}  &     \ding{53}  \\
		$ P $   & Period &  \ding{53}  &  \ding{53}  &     \ding{53}  \\
		$ r_s $   & $s$-orbital radius &  \ding{53}  &  \ding{53}     &  \ding{53}  \\
		$ E_{g,at} $ & HOMO-LUMO gap of atom &  \ding{53} &  \ding{53}  &     \ding{53}  \\
		$ E_{g,es} $ & Gap of elemental solid at  $\Gamma$-point&    & \ding{53}     &  \ding{53}  \\
		$ E_{f,es} $ & Energy of formation of elemental solid  &   &  \ding{53}     &  \ding{53}  \\
		$ E_{f,1di} $ & Energy of formation of mono-atomic dimer&    &     \ding{53} &       \ding{53}  \\
		$ E_{g,1di} $ & HOMO-LUMO gap of mono-atomic dimer  &      &  \ding{53}  &       \ding{53}\\
		\hline
		\multicolumn{5}{l}{Properties of two-atomic systems}\\
		\hline
		$ E_{f,bi} $ & Energy of formation of binary material&    &  \ding{53}     &  \ding{53}  \\
		$ E_{g,bi} $ & Gap  of binary material at the $\Gamma$-point    &  & \ding{53}      &  \ding{53}  \\
		$ d_{2di}  $ & Bond length of two-atomic dimer &    &  \ding{53}     &  \ding{53}  \\
		$ E_{f,2di} $ & Energy of formation of two-atomic dimer &    &  \ding{53}      &  \ding{53}  \\
		$ E_{g,2di} $ & HOMO-LUMO gap of two-atomic dimer &    &  \ding{53}    &     \ding{53}  \\
		\hline
		\multicolumn{5}{l}{Properties of tetrahedra}\\
		\hline
		$ d_{tet} $ 	& Bond length in tetrahedron  &    &    &    \ding{53}  \\
		$ E_{f,tet} $ 	& Energy of formation of tetrahedron &    &      &  \ding{53}  \\
		$ E_{g,tet} $ 	& HOMO-LUMO gap of tetrahedron        &    &     &  \ding{53}  \\
		\hline\hline
	\end{tabular}
\end{table}

Additional features are obtained by generalizations of equation \ref{eq:def_ave}, taking the $q$-th \emph{raw moments} of $ f_k $,
\begin{equation}
\bar{F}_{sk}^{q} = \frac{1}{m}\sum_{\gamma \in s} f_{k}(\gamma)^q
\label{eq:def_HM1}
\end{equation}
and the $q$-th root of the $q$-th \emph{central moments},
\begin{equation}
\widetilde{F}_{sk}^{q} = \left[ \frac{1}{m}\sum_{\gamma \in s} \left(f_{k}(\gamma) - F_{sk}\right)^q \right] ^{1/q}
\label{eq:def_HM2}
\end{equation}
where $q \in \mathbb{N}, \, q \geq 2 $ defines the order of the moment. Equation \ref{eq:def_HM1} defines an average of higher powers of the primary features. The central moments (equation \ref{eq:def_HM2}) carry information on the distribution of $f_k$ relative to the mean. The case $q=2$ can be understood as the standard deviation of $f_k$, and $q=3$ is related to its skewness. 

In many cases, materials' properties depend on differences in atomic properties, both with respect to geometry or composition. For instance, differences in atomic size can give rise to deviations from Vegard's law  \cite{DentonAshcroft1991,Murphy2010}. We address this fact by augmenting the pool of candidate features with averages over nearest neighbor differences of the atomic primary features:
\begin{equation}
\Delta F_{sk} \equiv \frac{1}{m} \sum_{\gamma = \langle i,j \rangle \in s} \left| f_{k}(i) - f_{k}(j) \right|.
\label{eq:def_next_neighbor_differences}
\end{equation}
Likewise, we also generalize this average by $q$-th raw  ($\bar{\Delta} F^{q}_{sk}$) and central ($\widetilde{\Delta} F^{q}_{sk}$) moments as in equations \ref{eq:def_HM1} and \ref{eq:def_HM2}. Primary features, $F_k$, that are derived from building blocks consisting of single crystal sites (\eg $f_k=Z$) will be independent to the atomic arrangement. On the contrary, if the averaging is carried out over building blocks containing more atoms, there will be an impact of the local atomic environment. An example of this are the tetrahedra, capturing the short-range order of five atomic sites (the center and the corners). Thus, the bond length (feature $f_k=d_{tet}$ in Table~\ref{tab:features_a}) may have different values dependent on the configuration of atoms in the tetrahedron.

To obtain the full feature space, we combine the primary features $F_k$ by mathematical operations in analogy to \cite{Ghiringhelli2015} and \cite{Ghiringhelli2017}:
\begin{enumerate}
\item First, the primary features $F_k$ obtained from equations \ref{eq:def_ave}, \ref{eq:def_HM1}, \ref{eq:def_HM2}, \ref{eq:def_next_neighbor_differences}, with $f_k$'s as listed in tables \ref{tab:features_a} and \ref{tab:features_e}, are added to the pool. These features are called tier 0, $\mathcal{T}_0$.
\item Tier 1, $\mathcal{T}_1$, comprises the features obtained by the following procedure:  First, the binary operations $(\cdot+\cdot)$ and $|\cdot-\cdot|$ (here $\cdot$ represents a feature) are applied to all combinations of features $F_k$ in $ \mathcal{T}_0 $, provided they have the same physical dimension. Then, the resulting expressions are acted on by the unary operators $(\cdot)^n$ ($n \in \left\lbrace \pm 1/3, \pm 1/2, \pm 2, \pm 3 \right\rbrace$), $\exp(\cdot)$, $1/\exp(\cdot)$, $\log(\cdot)$, and $1/\log(\cdot)$. The power $n=1$ is not included since it leads to linear combinations already represented by two-dimensional descriptors.
\item Furthermore, the previously generated features are combined by the product $(\cdot*\cdot)$ operation, yielding tier 2, $ \mathcal{T}_2 $. This can quickly lead to a combinatorial explosion of the number of features and thus become computationally untractable. As a bypass, we also define a restricted subspace $ \mathcal{T}_2^{\mathrm{r}} $ formed by all the products between one feature from $ \mathcal{T}_0 $ and one from $ \mathcal{T}_1 $. Note that all the features in $\mathcal{T}_{1}$ and $\mathcal{T}_{2}$ introduce non-linearities in the feature space. 

\end{enumerate}
The final feature space is then obtained from the union of all considered tiers. The corresponding sizes for the two learning tasks are displayed in table \ref{tab:size_of_feature_spaces}. We generally will use the symbols $ \mathcal{T}_i $ as a shorthand notation for feature spaces made up from tiers \emph{up to} $ \mathcal{T}_i $. 

\begin{table}[h]
\caption{Feature-space sizes, $N_\mathcal{F}$, used for the comparison of LASSO and SISSO as well as the model selection strategies for learning the lattice parameters (left columns) and the energy of mixing (right columns).}
\label{tab:size_of_feature_spaces}
\begin{center}
\begin{tabular}{|c|c|c|c|}
\hline 
\multicolumn{2}{|c|}{$a$} & \multicolumn{2}{c|}{$E_\mathrm{mix}$}\tabularnewline
\hline 
$ \mathcal{T}_i $ & $ N_\mathcal{F} $ & $ \mathcal{T}_i $ & $ N_\mathcal{F} $\tabularnewline
\hline 
\hline 
$ \mathcal{T}_1 $ & 5257 & $ \mathcal{T}_1 $ & 8736\tabularnewline
\hline 
$ \mathcal{T}_2^\mathrm{\,r} $ & 239797 & $ \mathcal{T}_2^\mathrm{\,r} $ & 572351\tabularnewline
\hline 
$ \mathcal{T}_2 $ & 13820653 & $ \mathcal{T}_2 $ & 38163216\tabularnewline
\hline 
\end{tabular}
\end{center}
\end{table}

\section{Model selection strategies} \label{sec:model_selection_strategies}

Applying a model-selection method to the total available data set, guarantees optimal interpolation within these data, it may, however, poorly perform with respect to predictions on new data. To evaluate the predictive power of the model, typically, cross-validation (CV) \cite{Friedmann2001} is used, partitioning the data into training set and test set, and building the model upon the training data only. Usually, this procedure is repeated several times on different training and test sets until convergence is reached. CV can, however, not only be used for model assessment but also for \emph{model selection}, \ie to choose between several competing candidate models. In this work, we devise strategies that transfer the concept of using CV to identifying descriptors as described in section \ref{sec:background}.

In the following, we describe three model selection strategies that we probe in this work. In the first one, which corresponds to a typical usage of CV, termed $\mathcal{S}_\mathrm{all} $, some kind of learning method $ \mathcal{L} $ is applied to \emph{all} data to determine an optimal model ${\cal D}$ by minimizing a suitable cost function, here the root mean square error (RMSE). In $ \mathcal{S}_\mathrm{all} $, $N_{\rm CV}$-fold cross-validation runs are performed in order to quantify the model's performance. Importantly, at each CV split $i=1$-$N_{\rm CV}$, $\mathcal{L}$ has to be applied to the training set $S_{\mathrm{tr}}^{(i)}$ in exactly the same way as before to the total data (\ie with the same hyperparameters defining the subset size $\tilde{M}$, the descriptor dimension $\Omega$, etc.). This procedure gives $N_{\rm CV}$ models ${\cal D}^{(i)}$ together with their individual training and test errors, $ \textrm{Err}_{\mathrm{tr},i} $ and $ \textrm{Err}_{\textrm{te},i} $, respectively. The frequency of CV splits $i$ in which the same model is found, ${\cal D}^{(i)}={\cal D}$, serves to measure the stability of ${\cal D}$ in \cite{Ghiringhelli2015} and \cite{Ghiringhelli2017}.

By contrast, our new strategies $\mathcal{S}_\mathrm{tr}^\mathrm{CV}$ and $\mathcal{S}_\mathrm{te}^\mathrm{CV}$ use the CV error, \ie the average errors $\overline{\mathrm{Err}}_\mathrm{tr}$ and $ \overline{\mathrm{Err}}_\mathrm{te} $ for model selection. Both strategies consist of the same three steps, (1) a pre-selection of candidate models $\lbrace \mathcal{D}^{(j)} \rbrace$, through a CV procedure; (2) the calculation of the average training and test errors, $\overline{\textrm{Err}}_{\mathrm{tr}}$ and $\overline{\textrm{Err}}_{\mathrm{te}}$, for these candidates via a second CV procedure, and (3) the selection of the best candidate from the $\lbrace \mathcal{D}^{(j)} \rbrace$ based on the average errors. The difference between the strategies is that the final selection of the best candidate model in step (3) is performed using average \emph{training} errors in $\mathcal{S}_\mathrm{tr}^\mathrm{CV} $ and average \emph{test} errors in $\mathcal{S}_\mathrm{te}^\mathrm{CV}$. Note that step (2) is carried out by a CV run where in each split individual training and test errors are calculated for \emph{all} candidates. In contrast to the CV procedure of step (1) or to the CV procedure in $\mathcal{S}_\mathrm{all}$, this step does not include any model selection but solely fits the free parameters $\mathbf c$ (see Eq.~\ref{eq:model}) for all candidates to the training set of the corresponding split. This serves to generate unbiased values for $ \overline{\textrm{Err}}_{\mathrm{tr}} $ and $ \overline{\textrm{Err}}_{\mathrm{te}} $ since, even if a model was selected in many CV-splits in step 1, it may still perform poorly for the splits in which it was not selected. 

Formally, step (3), yielding the candidate with the lowest average error, can be expressed as
\begin{equation}
\underset{\mathcal{D}^{(j)}}{\mathrm{argmin}}\,\overline{\textrm{Err}}_{\mathrm{tr}}(\mathcal{D}^{(j)})\,\,\,\,\,\textrm{or}\,\,\,\,\,\underset{\mathcal{D}^{(j)}}{\mathrm{argmin}}\,\overline{\textrm{Err}}_{\textrm{test}}(\mathcal{D}^{(j)})
\end{equation}
in $\mathcal{S}_\mathrm{tr}^\mathrm{CV}$ and $\mathcal{S}_\mathrm{te}^\mathrm{CV}$, respectively.

\section{Application to group-IV ternary alloys} \label{sec:application}

In the following, we demonstrate and compare the approaches, introduced in sections \ref{sec:construction_of_features} and \ref{sec:model_selection_strategies}, with the example of  ternary group-IV compounds in the zincblende structure as described in Section \ref{subsec:ternaries}. Since an important aim is to keep the dimension $\Omega$ of the descriptors low, we investigate $\Omega=1,\cdots,5$. The number of relevant features in LASSO+$\ell_0$ and SISSO is set to $\tilde{M}=30$ (see Sec.~\ref{subsec:compressed_sensing_methods}). For this value, an exact solution of the $\ell_0$ problem is still computationally feasible for $\Omega=5$. For SISSO, we use the selection operator SO($\ell_0$), \ie  the $\Omega$-dimensional descriptor is obtained by explicitly solving the $\ell_0$-problem on the actual union of subspaces. The latter is obtained by iterating SISSO five times, yielding in each iteration subspaces $\bm{S}_i$ of dimension 6, and then taking the union $\cup_{i=1}^{5} \bm{S}_i$, of dimension $\tilde{M}=30$. Unless indicated otherwise, we use the RMSE as an error metric.

As explained in Section \ref{subsec:ternaries}, the target properties $a$ and $\textrm{E}_\textrm{mix}$ are influenced by both the composition and the actual atomic arrangements at fixed composition. In order to quantify how well the different models capture them, the RMSE is employed. This measure, though, is not useful to assess a model's ability to capture the mere effect of atomic arrangements, as a model may capture them well but still poorly score on the RMSE. To this end, we first calculate the mean absolute error (MAE) between the centered property ($P-\langle P\rangle$) and the centered prediction ($\hat{P}-\langle \hat{P}\rangle$) at fixed composition, and then average this for all considered compositions:

\begin{equation}
\textrm{MAE}_{\textrm{arr}} \equiv 
\overline{
\Bigl< 
| (P - \langle P \rangle) -
(\hat{P} - \langle \hat{P} \rangle) | 
\Bigr>.
}
\label{eq:def_arrangement_error}
\end{equation}
Here, the overline indicates the average over all compositions $(x,y)$, and $\langle\cdot\rangle$ is the average over atomic arrangements within composition $(x,y)$. The subscript "arr" stands for "arrangement". This quantity is invariant under the addition of a constant to the predictions, either concerning the whole data set or the data with a certain composition. 

All the numerical computations using LASSO+$\ell_0$, SISSO, and the various feature selection strategies, are performed with an in-house Python code.

\subsection{Feature spaces} \label{subsec:feature_spaces_application}

In this work, we generate three different feature spaces. The corresponding primary features are listed in Tables \ref{tab:features_a} and \ref{tab:features_e}, respectively, for the two different learning tasks. They contain physical properties of single atoms, molecular dimers, tetrahedral clusters, as well as elemental and binary solids. The radii $r_s$, $r_p$ and $r_d$ are defined as the maximum radial probability density for the $s$-, $p$- and $d$-orbitals of the isolated atoms, respectively, taken from \cite{Ghiringhelli2017}. Dimer bond-lengths ($d_{1di}$, $d_{2di}$) and bulk lattice-constants ($a_{es}$, $a_{bi}$) are obtained by relaxation. The lattice constants of the elemental solids ($a_{es}$) are computed in the diamond structure, those of the binary materials ($a_{bi}$) in the zincblende structure.  We also consider all tetrahedral clusters present in the ternary materials. These are relaxed under the constraint of being fully symmetric, \ie determined by the distance $d_{tet}$ between the central atom and the corner atoms. Lattice constants, dimer bond lengths, and tetrahedral bond lengths are scaled to radius equivalents (\eg dimer lengths are divided by 2). This is done to allow for a better interpretation of features that algebraically combine these properties. Also energies of formation and Kohn-Sham band-gaps are included. For bulk materials, we consider the values at the $\Gamma$-point, while for finite systems, the differences between the highest occupied and the lowest unoccupied molecular orbitals (HOMO-LUMO gap). For the energies of formation, the ground-state energies of the isolated atoms are used as a reference. 

The feature spaces ($X_{a,i}$ and $X_{E,i}$, $ i=1,\ldots,3 $) are constructed for the two learning tasks using the primary features listed in Tables \ref{tab:features_a} and \ref{tab:features_e}, respectively. The first one, $X_{a/E,1}$, contains single-atom features only, \ie any average in equation \ref{eq:def_ave} will be over atomic building blocks only ($\gamma = i$). $X_{a/E,2}$ additionally contains pair features which are averaged over all $\gamma = \langle i,j \rangle$, and $X_{a/E,3}$ adds tetrahedral features.

\subsection{Complexity of feature space} \label{subsec:analysis_of_complexity}

In the following analysis, we choose the model-selection strategy $\mathcal{S}_\mathrm{all}$ in combination with $\textrm{LASSO}+\ell_{0}$. We verified that the use of $\textrm{SISSO}$ leads to similar results.

\subsubsection{Impact of mathematical operations.} \label{subsubsec:analysis_of_complexity_operations}

\begin{figure}[htb]
\centering
\includegraphics[width=0.8\textwidth]{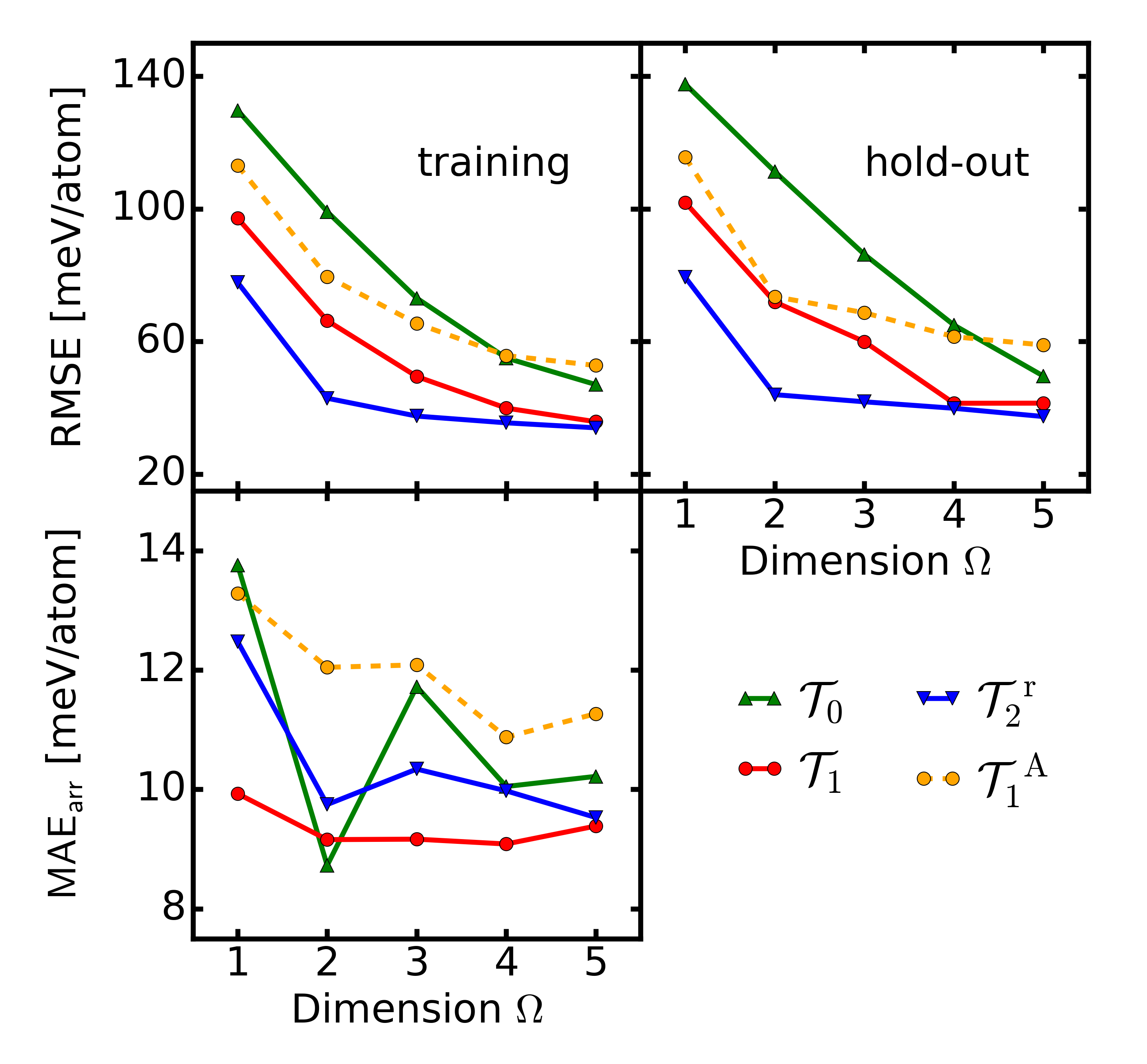}
\caption{Training errors (top left), hold-out errors (top right) and $ \mathrm{MAE}_\mathrm{arr} $ (bottom) for predicting $E_\textrm{mix}$ when mathematical operations of different complexity are employed. The superscript A indicates the exclusion of higher moments from $\mathcal{T}_1$. All feature spaces are based on  $X_{E,2}$.}
\label{fig:errors_complexity_operations_e}
\end{figure}

We start by studying how the mathematical operations used to construct the feature space impact the performance of the resulting descriptors, in terms of the tiers $\mathcal{T}_i,\,i=0,1,2$ (Section \ref{sec:construction_of_features}) as well as the specific operations. Thereby, we restrict the analysis to the learning of $E_{\mathrm{mix}}$ and note that the learning of the lattice constant leads to very similar results.

Figure \ref{fig:errors_complexity_operations_e} shows the RMSE of the training and the hold-out set as well as the $\textrm{MAE}_\textrm{arr}$ on the training data of the resulting descriptors for this learning task.  We use feature spaces based on single-atom and pair features ($X_{E,2}$) on the three complexity levels $\mathcal{T}_0$, $\mathcal{T}_1$ and $\mathcal{T}_2^\textrm{\,r}$ \cite{notet2r}. Furthermore, to evaluate the effect of the higher moments, we consider the feature space where the respective elements are eliminated from $ \mathcal{T}_1$ (indicated by $\mathcal{T}_1^{A}$) As expected, an increase of complexity overall reduces the fitting error. The decrease of the hold-out error indicates that the models don't suffer from overfitting. For $\mathcal{T}_1$ and $\mathcal{T}_2^\textrm{\,r}$, the training error converges to  about $\sim$30\,meV/atom. This means that \textit{optimal} models are obtained for $\mathcal{T}_1$ at $\Omega=5$ and for $\mathcal{T}_2^\textrm{\,r}$ already at $\Omega=3$. For $\mathcal{T}_0$, a higher-dimensional descriptor could reach even lower training errors. Arrangement effects are surprisingly well captured by medium complexity, \ie tier $\mathcal{T}_1$. Notably, for the 2D case, even $\mathcal{T}_0$ does best. However, the differences in $\textrm{MAE}_\textrm{arr}$ are generally rather small. Finally,  the worse performance of $\mathcal{T}_1^{A}$ as compared to $\mathcal{T}_1$, indicates that the higher moments accounting for non-linearities turn out to be an essential ingredient in feature construction.

\begin{figure}[htb]
\centering 
\includegraphics[width=0.7\textwidth]{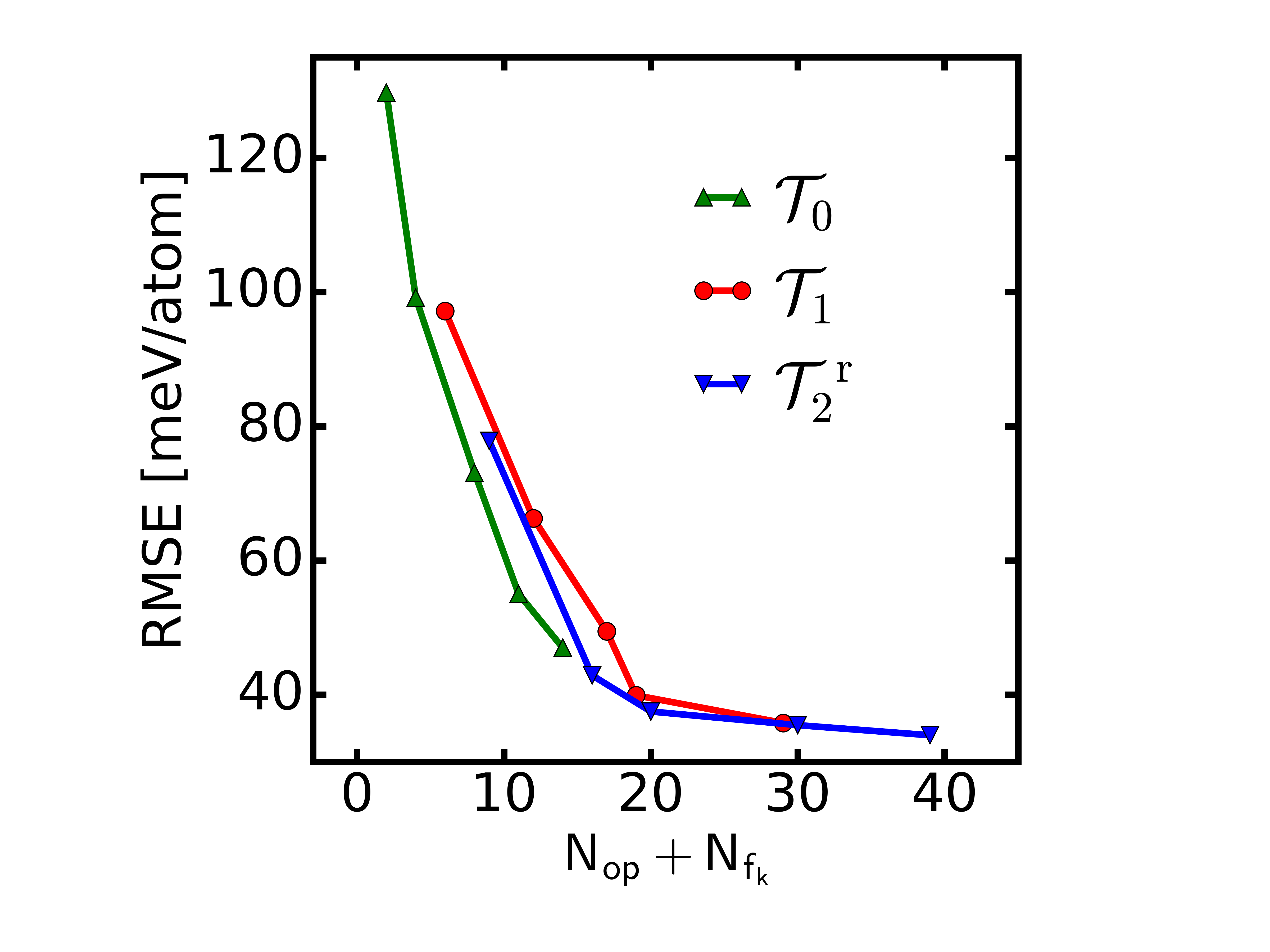}
\caption{Training error as a function of feature complexity, measured by $N_\textrm{op}+N_{f_{k}}$, for different tiers.}
\label{fig:feature_complexity}
\end{figure}

From the previous analysis, it becomes apparent that, rather independently of the tier used (or the feature space, see below), the increase in descriptor complexity improves the fit. To quantify this qualitative observation, we define a practical measure for the complexity as the sum of two terms, namely, the total number of operations in a descriptor, $N_\textrm{op}$, and the number of primary features, $N_{f_{k}}$. While the former captures the algebraic complexity, the latter accounts for the amount of physical information in the descriptor. In figure \ref{fig:feature_complexity}, the training errors are plotted versus this complexity measure. Interestingly, the curves of the three tiers approximately follow a common line. This uncovers a general dependency of the fitting error on the total complexity where the different tiers cover different regions of complexity.

Table \ref{tab:mathematical_complexity_1D_2D} summarizes the 1D and 2D descriptors found for the tiers $\mathcal{T}_0$, $\mathcal{T}_1$, and $\mathcal{T}_2^\textrm{\,r}$, along with the complexity measures for each of them. 
\begin{table}[htb]
\begin{center}
\caption{1D and 2D descriptors resulting from feature spaces with complexity $\mathcal{T}_0$, $\mathcal{T}_1$, and $\mathcal{T}_2^\textrm{\,r}$ together with the number of operations, $N_{\mathrm{op}}$, and the number of basic features, $ N_{f_{k}} $, considered in the descriptors.}
\label{tab:mathematical_complexity_1D_2D}
\begin{tabular}{|c|l|l|c|c|}
\hline 
$ \mathcal{T}_i $ & \multicolumn{2}{|c|}{1D} & $N_{\mathrm{op}}$ & $N_{f_{k}}$ \T\B \tabularnewline
\hline 
$ \mathcal{T}_0 $ & \multicolumn{2}{|l|}{$\tilde{E}_{g,es}^{2}$} & 1 & 1 \T\B \tabularnewline
\hline 
$ \mathcal{T}_1 $ & \multicolumn{2}{|l|}{$(\tilde{E}_{g,es}^{2} + \tilde{E}_{g,1di}^{2})^3$} & 4 & 2 \T\B \tabularnewline
\hline 
$ \mathcal{T}_2^\mathrm{\,r} $ & \multicolumn{2}{|l|}{$(\bar{E}_{f,1di}^{2}) \cdot \sqrt{\left| \bar{r}_s^{2} - \bar{d}_{2di}^{2} \right|}$} & 6 & 3 \T\B \tabularnewline
\hline 
\hline 
$ \mathcal{T}_i $ & \multicolumn{2}{|c|}{2D} & $N_{\mathrm{op}}$ & $N_{f_{k}}$ \T\B \tabularnewline
\hline 
$ \mathcal{T}_0 $ & $ \tilde{d}_{2di}^{2} $ &  $ E_{g,bi} $ & 2 & 2 \T\B \tabularnewline
\hline 
$ \mathcal{T}_1 $ & $(\tilde{E}_{g,es}^{2} + \tilde{E}_{g,2di}^{2})^3$ & $\exp( \Delta E_{g,es} + \Delta E_{g,1di} )$ & 9 & 3 \T\B \tabularnewline
\hline 
$ \mathcal{T}_2^\mathrm{\,r} $ & $ \tilde{d}_{2di}^{2} \cdot \left( \bar{\Delta} E_{g,at}^{2} + \bar{E}_{g,bi}^{2} \right)^2 $ & $ E_{m,es} \cdot \sqrt[3]{\left| r_s - d_{2di} \right|} $ & 11 & 5 \T\B \tabularnewline
\hline 
\end{tabular}
\end{center}
\end{table}

\begin{figure}[htb]
\includegraphics[width=\textwidth]{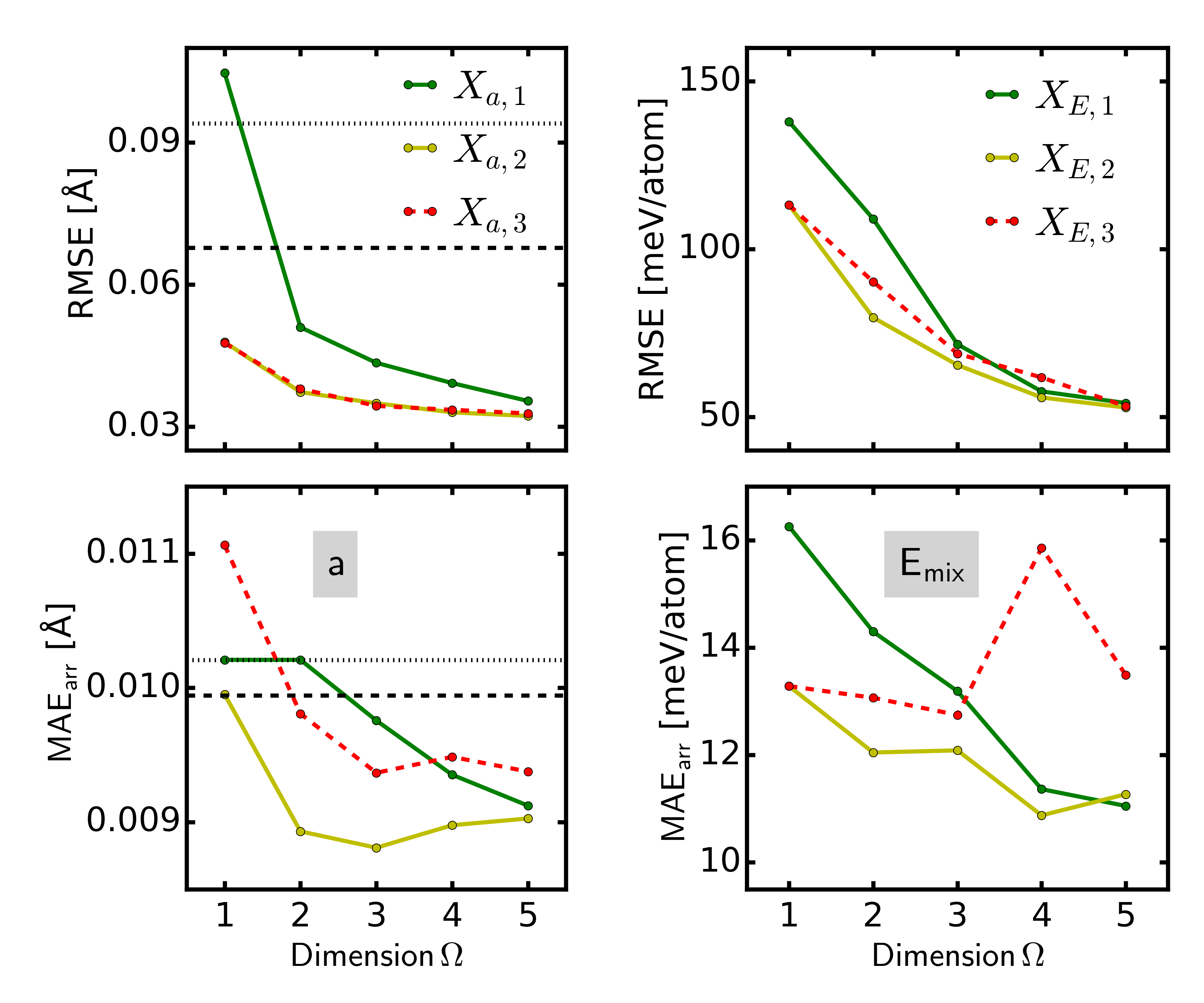}
\caption{Average training errors of different feature spaces $X$ for learning the lattice constant (left) and the energy of mixing (right). The top panels show the RMSE versus descriptor dimension $\Omega$, the bottom panels the $\mathrm{MAE}_{\mathrm{arr}}$. Black dotted lines: error of interpolation between pristine materials (Vegard's law), black dashed lines: error of interpolation between binary materials.}
\label{fig:errors_complexity_features}
\end{figure}

\subsubsection{Influence of primary features.} \label{subsubsec:analysis_of_complexity_properties}
We now study how the adoption of different feature spaces influences the obtained descriptors. To better emphasize the mere effect of the feature space on the results, we limit this analysis to the low-complexity tier $\mathcal{T}_1^{A}$ defined in the previous section. The left panels of figure \ref{fig:errors_complexity_features} show the RMSE and the $\textrm{MAE}_{\textrm{arr}}$ for the training data versus descriptor dimension $\Omega$ for predicting lattice constants for the three feature spaces $X_{a,1-3}$.   The dotted and dashed horizontal lines indicate the errors of Vegard's law and the averaged binary lattice constant (equation \ref{eq:def_ave}), respectively. Depending on the composition only, the former is totally blind to effects of atomic arrangement. The latter captures them to some extent, thus exhibiting a somewhat smaller error. For all three feature spaces, the increase in dimension $\Omega$, naturally, lowers the training error, which is most distinctive for $X_{a,1}$. Except for the 1D descriptor of $X_{a,1}$ ($\mathbf{D}_\textrm{1D} = 1 / \sqrt[3]{r_p}$), all models outperform Vegard's law as well as the simple averaging procedure. Including pair quantities in the feature spaces, reduces the RMSE drastically. Adding also the tetrahedral features, by contrast, has a negligible effect on the training error, though the descriptors emerging from $X_{a,3}$ are not identical to those from $X_{a,2}$ but contain also the tetrahedral feature $d_{tet}$.
 
The $\textrm{MAE}_{\textrm{arr}}$, in the the lower panel, also shows an overall decrease with higher $\Omega$, but exhibits a  shallow minimum at $\Omega = 3$ for  $X_{a,2}$ and $X_{a,3}$. Even descriptors from the pure atomic feature space ($X_{a,1}$) can outperform Vegard's law. This is the case for $\Omega \geq 3$ where the descriptors include nearest-neighbor differences in the fashion of equation \ref{eq:def_next_neighbor_differences}.  The best result is achieved by combined atomic and pair features ($X_{a,2}$) for all descriptor dimensions considered. 
Surprisingly, $X_{a,3}$, though doing well in the training error, is performing worse in the MAE$_{\textrm{arr}}$. This is presumably due to the  rigidity of the tetrahedra, departing considerably from the actual physical situation in the solid. To sum up, optimal feature spaces for predicting lattice constants of the considered data, should be based on a combination of atomic and pair features, and the tetrahedra-based features can be discarded here safely.

The right part of figure \ref{fig:errors_complexity_features} shows RMSE and $\textrm{MAE}_{\textrm{arr}}$ for learning $E_{\mathrm{mix}}$. As expected, also in this case the fitting error decreases with increasing $\Omega$ for all $X_{E,i}$. Also expected, the 1D and 2D descriptors based on $X_{E,1}$ exhibit a larger RMSE than the ones from $X_{E,2}$ and $X_{E,3}$, while considering more complex descriptors ($\Omega \geq 3$) does not improve the RMSE anymore,\ie all three features spaces have similar error. Adding tetrahedral features, overall slightly increases the RMSE as compared to $X_{E,2}$, except for $\Omega = 1$ where $X_{E,2}$ and $X_{E,3}$ give the same descriptor. One would expect that offering more features cannot increase the error, and just the solution from the smaller feature space would be found. While this is true for solving the $\ell_0$ problem in an exact manner, for the approximation by $\mathrm{LASSO}+\ell_0$ this does not need to be the case \cite{lassol0}. 

Finally, we have a look at the related MAE$_{arr}$ (lower panel). In short, also for the energy of mixing, next-neighbor differences play an important role, where descriptors from $X_{E,2}$ generally perform best. Like for the RMSE, addition of tetrahedral features never leads to an improvement, most likely because the local environment in the real data often differs markedly from the regular tetrahedral shape. In contrast, the most basic atomic features are performing quite well.
 
\begin{figure}[htb]
\includegraphics[width=1.\textwidth]{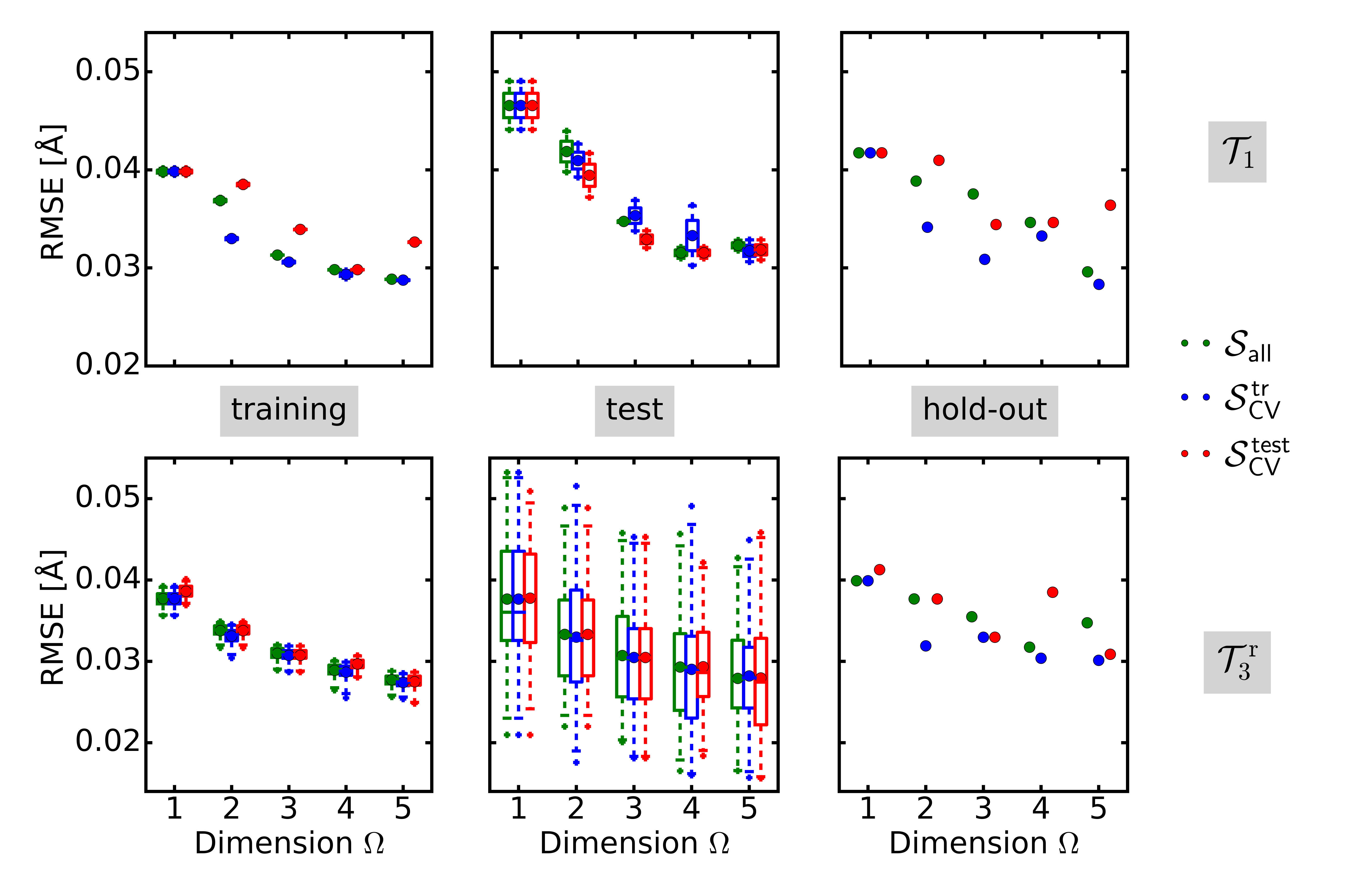}
\label{fig:compare_strategies_a_t2}
\caption{Training (left), test (middle), and hold-out (right )errors vs. descriptor dimension $\Omega$ for the different model selection strategies for predicting the lattice constant $a$. The top (bottom) plots refer to feature space complexity $\mathcal{T}_1 $ ($\mathcal{T}_2^\mathrm{\,r}$). The left and middle plots stem from  errors from 50 L($10\%$)OCV runs with fixed descriptor components. In the box plots, filled circles indicate the average errors, the borders of the boxes mark the 25th and the 75th percentile, and the middle line indicates the median. Whiskers mark the region between the 1st and the 99th percentile, and crosses indicate the minimum and the maximum of the distribution. The results for Vegard's law are independent of the dimension and all have a RMSE of about 0.07 \AA.}
\label{fig:compare_strategies_a_errors}
\end{figure}

\subsection{Model selection strategies} \label{subsec:comparison_model_selection_strategies}

Now we apply the different model selection strategies described in Section \ref{sec:model_selection_strategies}, to the prediction of lattice constants, employing $\mathrm{LASSO}+\ell_0$ and the feature spaces $\mathcal{T}_1$ and $\mathcal{T}_2^\textrm{\,r}$. The results depicted in figure \ref{fig:compare_strategies_a_errors}, comprise the distribution of training (left), test (middle) and hold-out errors (right). The error distributions for training and test data result from 50 CV runs where the descriptors obtained by the three strategies were kept fixed.

For all three strategies, the training errors don't show significant spread.  $\mathcal{S}^\textrm{tr}_\textrm{CV}$ exhibits slightly smaller values compared to the other two. For $\mathcal{T}_2^\textrm{\,r}$, $\mathcal{S}^\textrm{test}_\textrm{CV}$ is very similar as well, for $\mathcal{T}_1$ and some descriptor dimensions $\Omega$ it is markedly worse, however.  The test errors have a narrow distribution at smaller matrix size ($\mathcal{T}_1$, top panel) and a wide distribution at larger matrix size ($\mathcal{T}_2^\textrm{\,r}$, bottom panel). This is to be expected for increasing model complexity \cite{Friedmann2001}.  $\mathcal{S}^\textrm{test}_\textrm{CV}$ here performs better or similar to the other strategies, as expected, though no large difference between the strategies is apparent. Generally, $\mathcal{S}^\textrm{tr}_\textrm{CV}$ yields the smallest hold-out errors and $\mathcal{S}^\textrm{test}_\textrm{CV}$ tends to yield the largest. The results for learning $E_\textrm{mix}$ exhibit the analogous trends.

\subsection{LASSO versus SISSO} \label{subsec:LASSO_vs_SISSO}

\begin{figure}[htb]
\centering
\includegraphics[width=1.\textwidth]{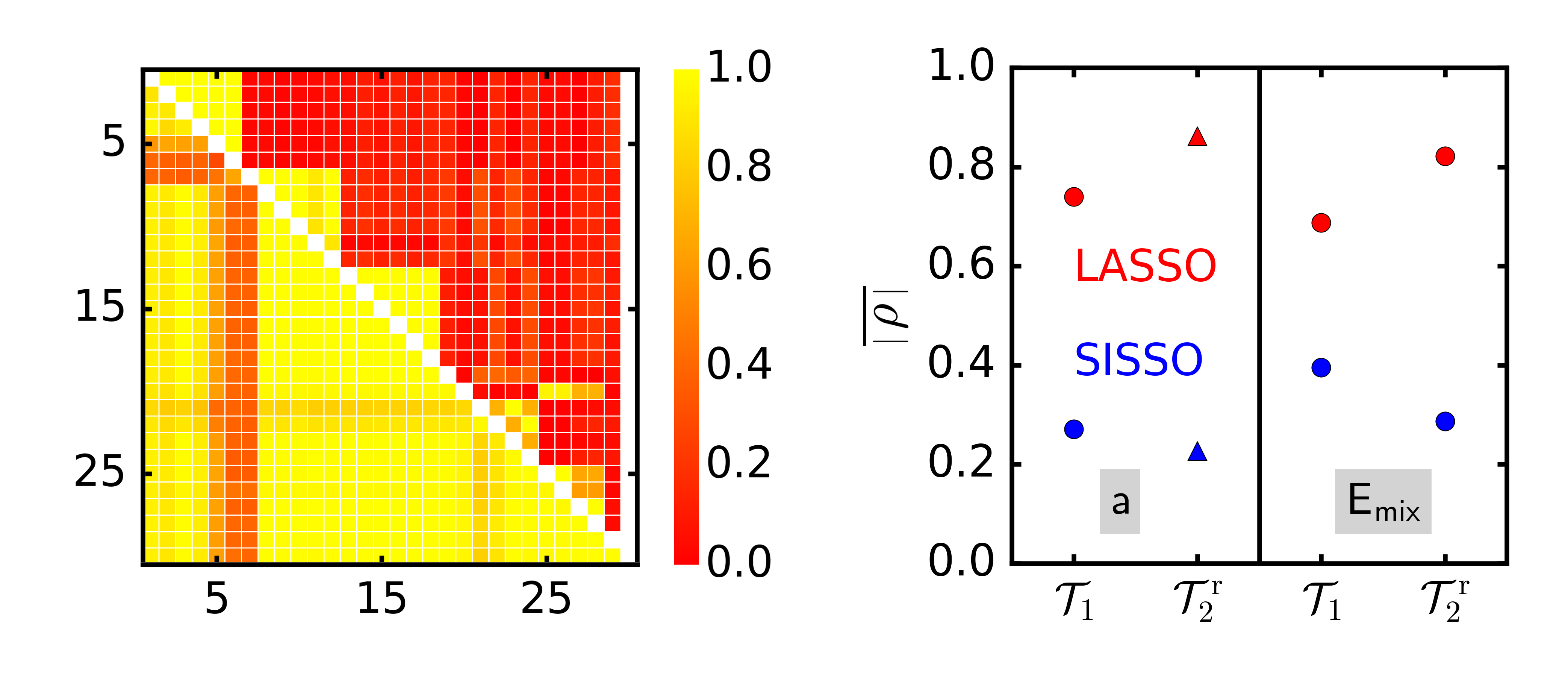}
\caption{ Left: Correlogram of the preselected subsets for learning lattice parameters with feature space complexity $\mathcal{T}_2^{\mathrm{\,r}}$ obtained from LASSO (lower triangle) and SISSO (upper triangle). The color code indicates the absolute Pearson correlation between 0 (red) and 1 (yellow). Right: Absolute Pearson correlation averaged over pre-selected subsets versus feature space complexity for predicting $a$ (left) and $E_\mathrm{mix}$ (right). The triangles indicate the cases analyzed in the left panel.}
\label{fig:compare_preselection}
\end{figure}

As described in Section \ref{subsec:compressed_sensing_methods}, SISSO has been proven to overcome the issue of strongly correlated features in $\mathrm{LASSO}+\ell_0$ \cite{Ouyang2018}, besides being able to deal with much larger feature matrices $\mathbf{D}$. Hence, it is interesting to compare these two feature selection methods for identifying optimal descriptors. To do so, we first compare the pre-selected subspaces and then the resulting 5D descriptors, applying the strategy $\mathcal{S}_\textrm{all}$. As a measure of (linear) correlations between features, we use $\left| \rho_{i,j} \right|$, the absolute value of the Pearson correlation coefficient. The symmetric matrix $\left| \rho_{i,j} \right|$ is visualized in figure \ref{fig:compare_preselection} in terms of a correlogram. Here, pairwise correlations of the pre-selected subspaces are shown for predicting the lattice constant, with LASSO in the lower-left triangle and SISSO in the upper-right one. (Note that the subset obtained with SISSO (LASSO) contains 29 (30) features. Such a difference may occur when certain features are repeatedly selected to different subsets in SISSO.) The feature space complexity here is $\mathcal{T}_2^{\mathrm{\,r}}$. It is a challenging case for LASSO where indeed many highly correlated candidate features dominate its subspace. By contrast, most of the features selected by SISSO are less correlated (predominance of the red color). The small yellow triangles inside the upper-right triangle correspond to the bunch-like subspaces which, by construction of SISSO, contain highly correlated features.

In the right panel of the figure, the mean absolute correlations $\overline{\left| \rho_{i,j} \right|}$ are compared for feature spaces of different complexity, for both predicting $a$ and $E_\mathrm{mix}$. Clearly, in all cases the overall correlation is significantly higher for subsets obtained with LASSO compared to SISSO. Moreover, at increasing matrix size, it increases for LASSO and decreases for SISSO. For the more difficult learning task of predicting $E_\mathrm{mix}$, the overall correlation is lower compared to the simpler task of learning $a$ when using LASSO, while for SISSO it is the opposite. As a general -- and expected -- conclusion, SISSO selects subsets with less redundancy, in particular for large and highly correlated feature spaces. This is a prerequisite for the following $\ell_0$ step to select low-correlated descriptors.

\begin{table}[]
\caption{Components of the 5D descriptors for predicting $a$ with feature space complexity $\mathcal{T}_2^{\mathrm{\,r}}$, obtained by applying an exhaustive $\ell_0$ search on the subsets from LASSO or SISSO, respectively. The components are sorted by their absolute Pearson correlation with $ a $ in decreasing order.}
\label{tab:components_5D_a_t3p}
\begin{center}
\begin{tabular}{|c|c|c|c|c|c|}
\hline
Component   & 1 & 2 & 3  & 4 & 5 \TTT\BBB \tabularnewline
\hline
LASSO 
& $ \frac{\bar{a}_{bi}^{2}}{\sqrt{d_{1di}}} $ 
& $ \frac{\bar{a}_{bi}^{2}}{a_{es}} $ 
& $ \sqrt[3]{\left|\bar{r}_p^{2}-\bar{d}_{2di}^{2}\right|} $ 
& $ \bar{d}_{2di}^3 \cdot \sqrt[3]{\left| \tilde{r}_d^{2}-\tilde{d}_{2di}^{3}\right|} $ 
& $ \frac{\tilde{d}_{2di}^{3}}{\sqrt[3]{\tilde{r}_s^{2}+\tilde{r}_s^{3}}} $ \TTT\BBB \tabularnewline 
\hline
SISSO 
& $ \frac{\bar{a}_{bi}^{2}}{\sqrt{\bar{r}_p^{2}+\bar{a}_{bi}^{2}}} $ 
& $ \frac{\tilde{Z}^{3}}{\bar{r}_p^{2}+\bar{r}_p^{3}} $ 
& $ \bar{r}_d^2\cdot\sqrt{\tilde{a}_{bi}^{3}} $ 
& $ \tilde{a}_{bi}^2\cdot(\tilde{d}_{1di}^2-\tilde{a}_{bi}^3)^3 $ 
& $ \frac{\tilde{P}^{3}}{\tilde{r}_s^{2}+\tilde{r}_d^{3}} $ \TTT\BBB \tabularnewline 
\hline
\end{tabular}
\end{center}
\end{table}

\begin{figure}[htb!]
\includegraphics[width=1.\textwidth]{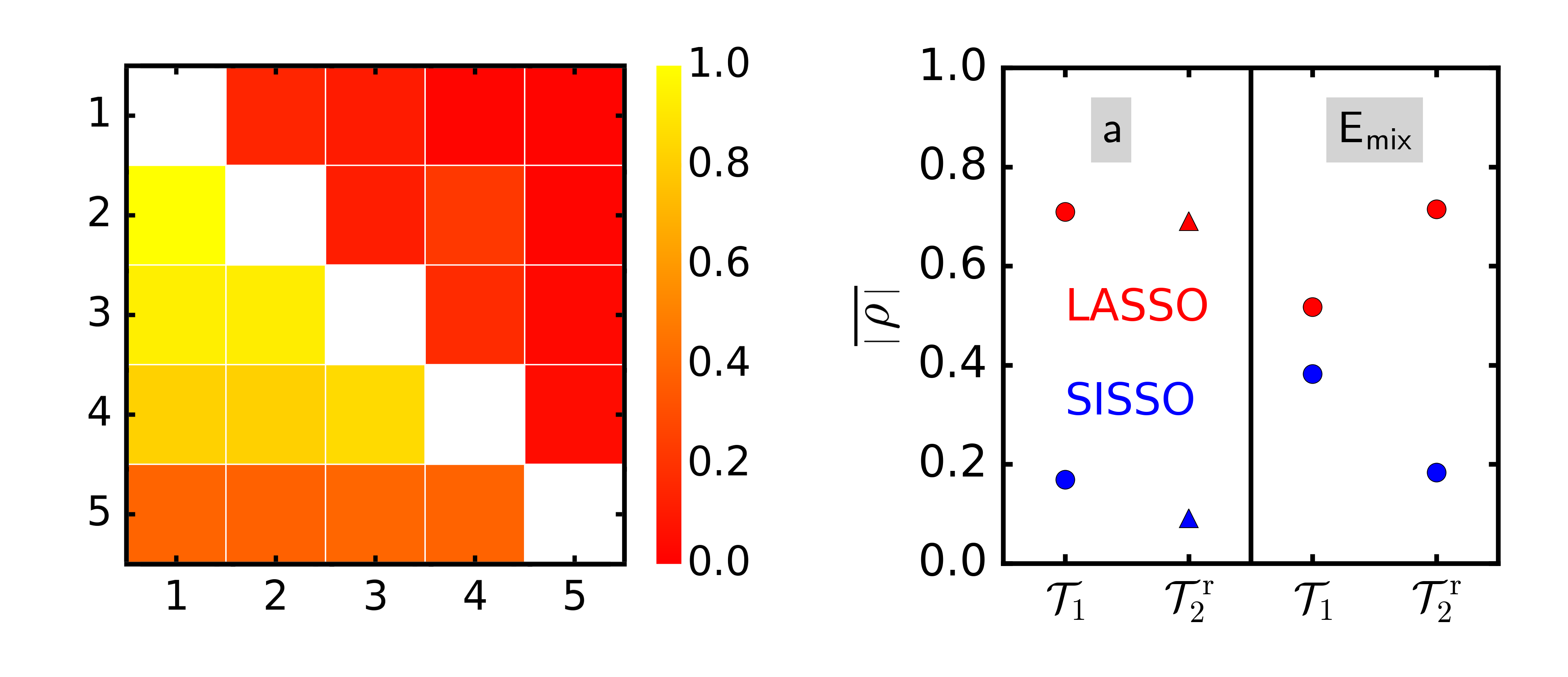}
\includegraphics[width=1.\textwidth]{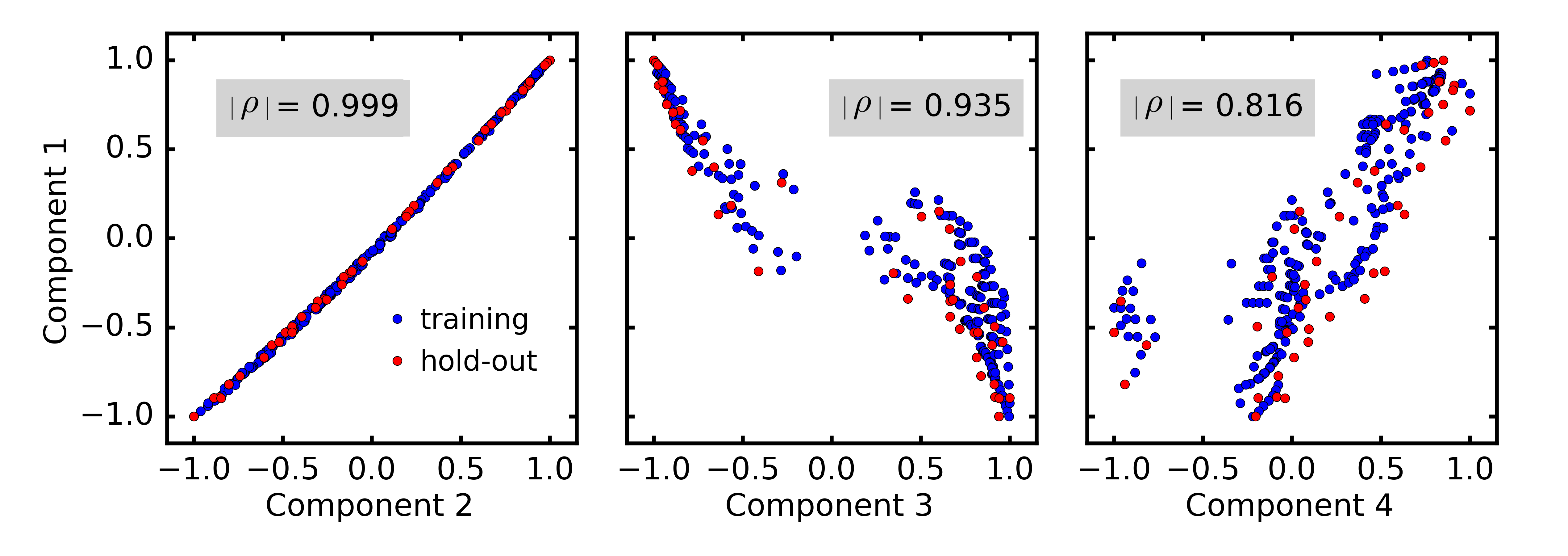}
\caption{Comparison of 5D descriptors obtained by LASSO and SISSO. Top left: correlogram of their components for learning $a$ with feature-space complexity $\mathcal{T}_2^{\mathrm{\,r}}$ (lower triangle: LASSO+$\ell_0$, upper triangle: SISSO+$\ell_0$). Top right: average Pearson correlation versus feature-space complexity for predicting $a$ and $E_\mathrm{mix}$.  The triangles indicate the cases analyzed in the left panel. Bottom: correlations between different components of the solution obtained by LASSO+$\ell_0$ on $\mathcal{T}_2^{\mathrm{\,r}}$ for predicting $a$. These pairs give examples of high, medium, and low correlation.}
\label{fig:compare_5D}
\end{figure}

For learning the lattice parameter at complexity level $\mathcal{T}_2^{\mathrm{\,r}}$, Table \ref{tab:components_5D_a_t3p} lists the components of the 5D descriptors obtained when applying the $\ell_0$ step to either the LASSO or the SISSO subspace. Figure \ref{fig:compare_5D} shows their correlations in the upper left panel. Since the preselected subspace offers only four low correlated pairs (see lower triangle in the left panel of figure \ref{fig:compare_preselection}), all possible 5D descriptors from the LASSO subspace inevitably contain at least one highly correlated feature pair. In fact, components 1 and 2 are particularly highly correlated ($\left| \rho_{1,2} \right| = 0.999$) and also appear to have a similar algebraic form (see table \ref{tab:components_5D_a_t3p}).  Also the correlations between components 1 and 3 as well as 2 and 3 are very high ($\left| \rho_{1,3} \right| = 0.935$ and $\left| \rho_{2,3} \right| = 0.929$, respectively), however, they seem to carry different information.  This can be seen in the bottom panel of the figure where the correlations are displayed for the training data and the hold-out set, exhibiting a larger spread (see middle panel). The right panel shows the correlations between components 1 and 4 ($\left| \rho_{2,3} \right| = 0.816$). By contrast to the LASSO descriptor, the 5D descriptor from SISSO does not contain highly correlated components. 

The upper right panel of the figure compares the mean correlation of the 5D descriptors, obtained by the two methods. For LASSO, increasing feature-space complexity rises $\overline{\left| \rho_{i,j} \right|}$  in case of $E_\mathrm{mix}$ and slightly drops in case of $a$. For SISSO, we observe a decrease in both cases, and the values are always lower than the LASSO counterparts.

\begin{figure}[h]
\centering\large \includegraphics[width=1.\textwidth]{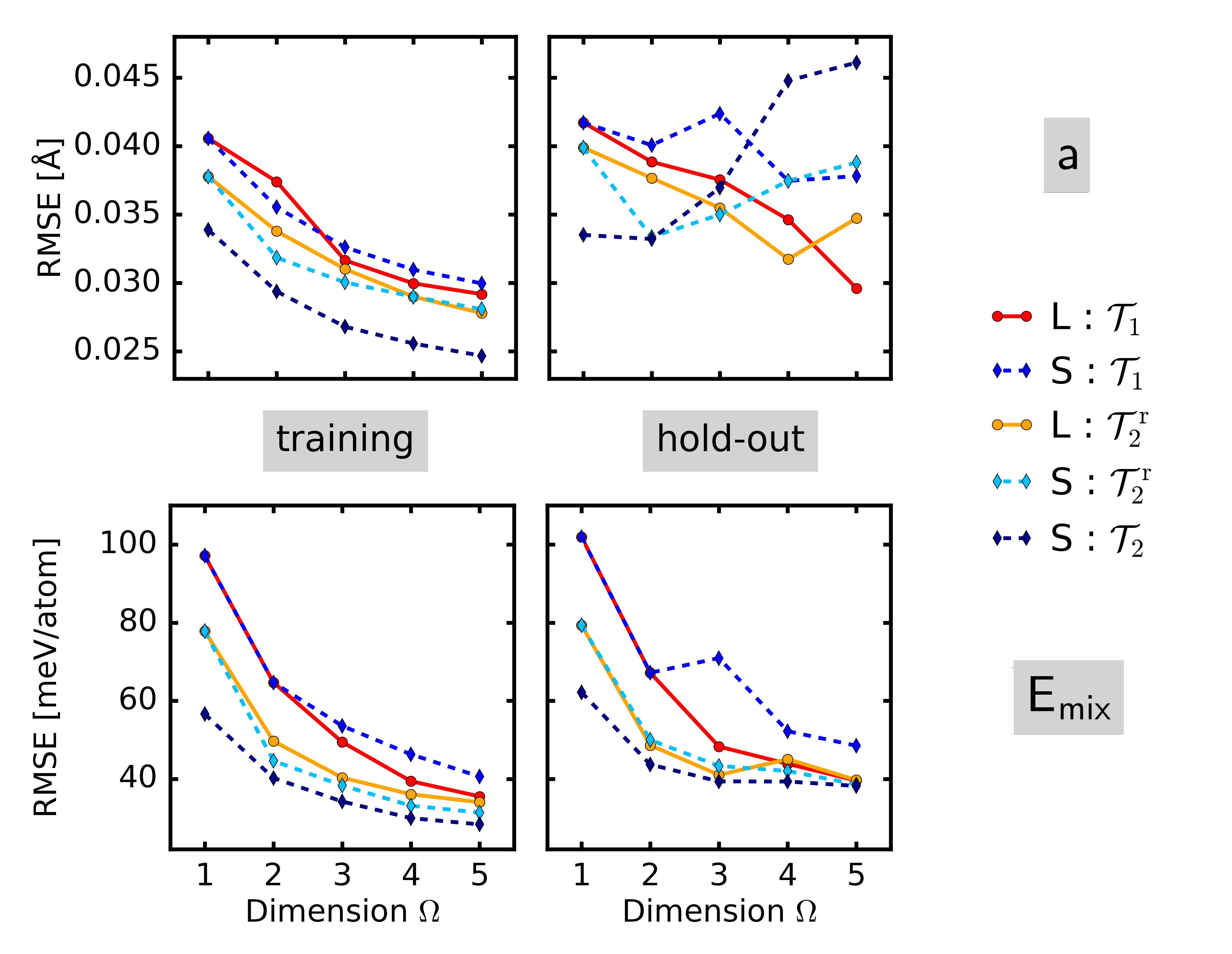}
\caption{Fitting and hold-out errors of descriptors up to $\Omega=5$ from LASSO (L) and SISSO (S) for various levels of feature-space complexity for predicting $a$ (upper panels) and $E_\mathrm{mix}$ (lower panels).}
\label{fig:errors_L_vs_S}
\end{figure} 

It is also interesting to compare the predictive power of the descriptors obtained from LASSO and SISSO. This is shown in figure \ref{fig:errors_L_vs_S}, displaying the fitting and hold-out errors for the two learning tasks. Thereby, feature spaces with complexity $\mathcal{T}_1$ and $\mathcal{T}_2^{\mathrm{\,r}}$ are considered. While the fitting errors do not differ much between LASSO and SISSO for both learning tasks, the hold-out errors for learning the lattice parameter (left panel) indicate that for larger $\Omega$, the SISSO descriptors may be overfitted. Additionally, results for $\mathcal{T}_2$ are shown obtained from SISSO only (LASSO being limited due to its high demands in terms of memory). This further reduces the fitting errors but enhances overfitting for $\Omega \geq 3$. In contrast, for learning $E_\mathrm{mix}$, there is no evidence of overfitting, even for $\Omega=5$. Also in this case, the higher complexity of feature space ($\mathcal{T}_2$) reduces the fitting error without leading to overfitting.

Overall, in the studied cases, the descriptors from LASSO and SISSO are very similar with respect to predictive power. As expected, descriptors from SISSO benefit from less redundant components and can handle big feature matrices.

\subsection{Optimal descriptors} \label{subsec:optimal_descriptors}

\begin{figure}[h]
\includegraphics[width=1.\textwidth]{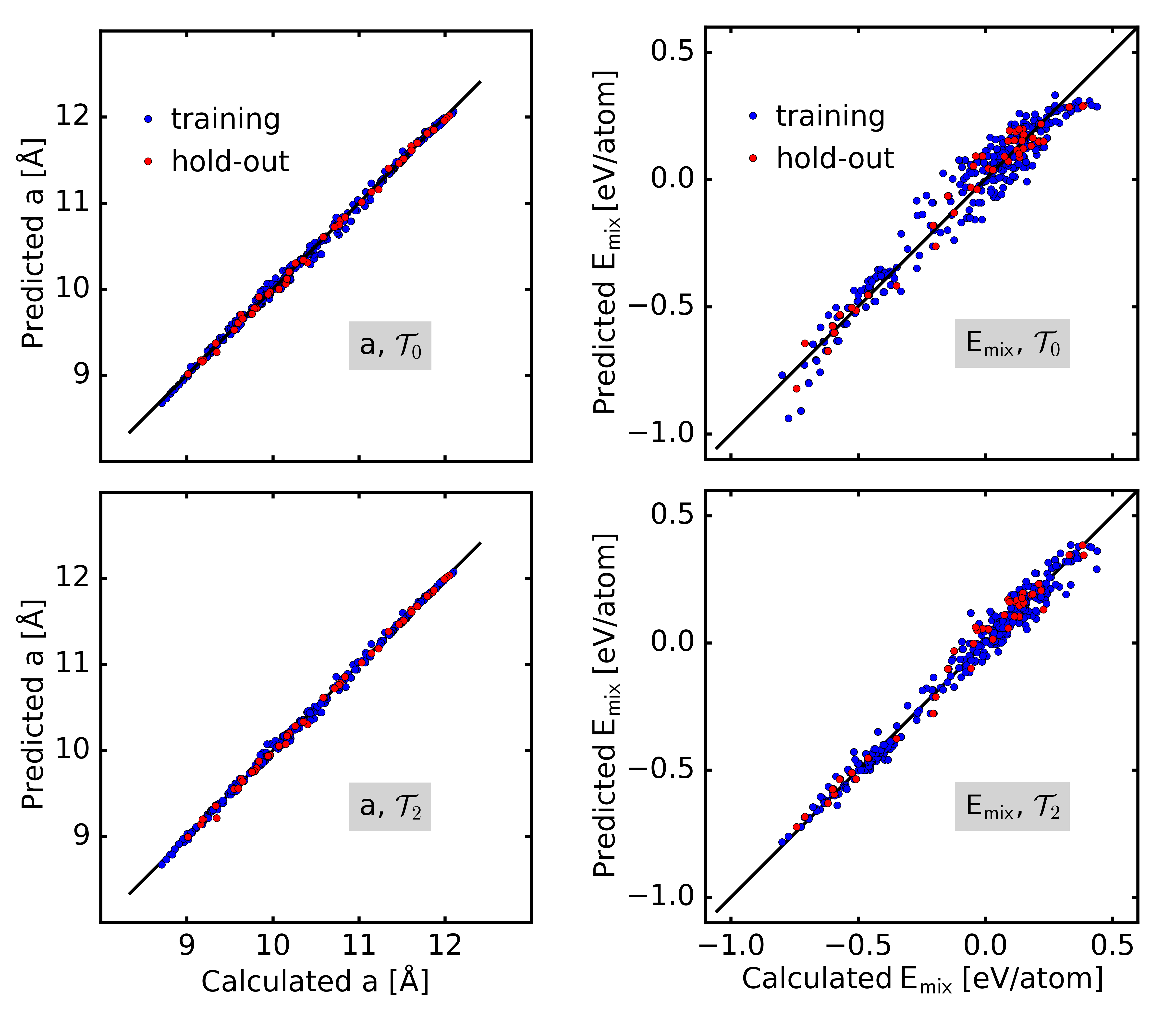}
\caption{Predictions by the optimal descriptors versus calculated values. Results for learning of $a$ are shown on the left, for $E_\mathrm{mix}$ on the right. In the two left plots, feature space complexity $\mathcal{T}_0$ was used, in the two right ones complexity $\mathcal{T}_2$.} 
\label{fig:corr_plots_optimal_descriptors}
\end{figure}

There is always some degree of arbitrariness as there are plenty of ways to construct candidate features. This also holds for model parameters like the descriptor dimension $\Omega$.  One would, for example, select a small value of $\Omega$ if simplicity is the main goal, at the expense of accuracy.  Moreover, when feature complexity increases, the selection methods yield more and more competing models of virtually the same predictive power -- which itself can be defined in different ways.  Thus, in table \ref{tab:optimal_descriptors_formulas}, we present some exemplary descriptors for both $a$ and $E_\mathrm{mix}$, based on the previous analyses. For $a$, we have selected the 3D descriptor on the very simple complexity level $\mathcal{T}_0$ and the 2D descriptor on the rather complex $\mathcal{T}_2$; for $E_\mathrm{mix}$, the 4D descriptor on $\mathcal{T}_0$ and the 2D descriptor on $\mathcal{T}_2$. All of these were identified by SISSO using the strategy $\mathcal{S}_\textrm{all}$. The corresponding error analysis is presented in table \ref{tab:performance_optimal_descriptors}. Figure \ref{fig:corr_plots_optimal_descriptors} visualizes the performance of the obtained models.

\begin{table}[h]
\caption{Components of the descriptors for predicting $a$ and $E_\textrm{mix}$ on complexity levels $\mathcal{T}_0$ and $\mathcal{T}_2$. The parenthesis indicate components that, when exchanged by each other, are virtually indistinguishable in performance.}
\label{tab:optimal_descriptors_formulas}
\begin{center}
\begin{tabular}{c|c|c}
Target & Complexity & Descriptor components\TTT\BBB \tabularnewline
\hline
\multirow{2}{*}{$a$}& $\mathcal{T}_0$
 & 
$a_{bi}$\hspace{5mm} 
$\tilde{a}_{bi}^{2}(\tilde{d}_{2di}^{2})$\hspace{5mm} 
$\tilde{a}_{bi}^{3}$ 
\T\B\tabularnewline
\cline{2-3}
& $\mathcal{T}_2$ & 
$\frac{\sqrt{\bar{a}_{es}^{2}+\bar{a}_{bi}^{2}}}{\sqrt[3]{\bar{r}_p^{2}+\bar{d}_{1di}^{2}}}$\hspace{5mm}
$\frac{\sqrt[3]{\left|\tilde{r}_d^{2}-\tilde{d}_{2di}^{3}\right|}}{(r_d+\tilde{r}_d^{3})^{3}}$
\TTT\BBB \tabularnewline
\hline
\multirow{2}{*}{$E_\mathrm{mix}$}& $\mathcal{T}_0$&
$\bar{Z}^{3}$\hspace{5mm}
$\tilde{P}^{2}$\hspace{5mm}
$\tilde{E}_{g,bi}^{2}$\hspace{5mm}
$\tilde{a}_{bi}^{3} \,(\tilde{d}_{2di}^{3})$
\T\B\tabularnewline
\cline{2-3}
&$\mathcal{T}_2$ &
$\frac{1}{E_{f,1di}+\tilde{E}_{g,at}^{2}}\cdot\frac{1}{\sqrt{E_{f,1di}+E_{g,at}}}$\hspace{5mm} 
$(E_{g,1di}+E_{g,bi})^2\cdot\sqrt{\left|\bar{r}_s^{3}-\bar{d}_{2di}^{3}\right|}$
\TTT\BBB \tabularnewline
\end{tabular}
\end{center}
\end{table}
 
\begin{table}[h]
\caption{Performance of the \textit{optimal} descriptors for predicting $a$ and $E_\textrm{mix}$. RMSE and MAE averaged over training and test sets of CV; MAE and maximum absolute error (maxAE) on the hold-out set; and MAE$_{arr}$. The mean absolute percentage error (MAPE) is shown in training and hold-out sets for learning $a$. Lenghts are in units of m\AA\ and energies in meV/atom.}
\label{tab:performance_optimal_descriptors}
\begin{center}
\setlength{\tabcolsep}{8pt}
\renewcommand{\arraystretch}{1.5}
\begin{tabular}{cc|ccc|c|c|c|c|c|c|c|c|c}
\multirow{2}{*}{\rot{Target}} & \multirow{2}{*}{\rot{Complexity}} & \multirow{2}{*}{$\Omega$} & \multirow{2}{*}{$N_{f_k}$} & \multirow{2}{*}{$N_\mathrm{op}$} & \multicolumn{3}{c|}{Training} & \multicolumn{2}{c|}{Test} & \multicolumn{3}{c|}{Hold-out} & \tabularnewline
 &  &  &  &  & \rot{RMSE} & \rot{MAE} & \rot{MAPE} & \rot{RMSE} & \rot{MAE} & \rot{MAE} & \rot{MAPE} & \rot{maxAE} & \rot{$\mathrm{MAE}_\mathrm{arr}$}\tabularnewline
\hline 
\hline 
\multirow{2}{*}{$a$} & $\mathcal{T}_0$ & 3 & 1 & 4 & 38 & 26 & 0.26 & 41 & 28 & 29 & 0.28 & 93 & 9\T\B\tabularnewline
 & $\mathcal{T}_2$ & 2 & 6 & 18 & 29 & 19 & 0.19 & 30 & 20 & 21 & 0.21 & 133 & 9\T\B\tabularnewline
\hline 
\multirow{2}{*}{$E_\textrm{mix}$} & $\mathcal{T}_0$ & 4 & 4 & 7 & 64 & 51 &-& 64 & 51 & 46 & - & 130 & 11\T\B\tabularnewline
 & $\mathcal{T}_2$ & 2 & 6 & 15 & 40 & 30 & - & 40 & 31 & 35 & - & 99 & 10\T\B\tabularnewline
\end{tabular}
\end{center}
\end{table}

We start with the 3D descriptor for $a$ on complexity level $\mathcal{T}_0$. It is favorable because it only contains the binary bulk lattice constant $a_{bi}$ and takes a simple form in the style of a polynomial expansion and thus effectively augments a pure interpolation between the binary materials. We note that the three components are not redundant as they do not exhibit a high pairwise correlation (maximum $\left| \rho_{i,j} \right|=0.49$).

The alternative descriptor presented for this task, the 2D descriptor on $\mathcal{T}_2$, has much higher informational ($N_\mathcal{F}=6$) and algebraic ($N_\mathrm{op}=18$) complexity as the former ($N_\mathcal{F}=1$ and $N_\mathrm{op}=4$). It uses all basic properties of length dimension from our pool of basic properties $\lbrace f_k \rbrace$, except $r_s$. The increased complexity results in an improved predictive performance on the expense of a simple, intuitive interpretation. Note also here that the two components are hardly correlated ($\left| \rho_{1,2} \right|=0.14$). As can be seen from figure \ref{fig:corr_plots_optimal_descriptors}, both presented descriptors for $a$ have good predictive power without any outliers. So, apart from the fact that the descriptor obtained from complexity $\mathcal{T}_2$ exhibits slightly smaller errors, it appears sufficient to stay with that from $\mathcal{T}_0$.

$\tilde{E}^2_{g,bi}$ and $\tilde{a}^3_{bi}$ appear as the third and fourth component for learning $E_\mathrm{mix}$. As the lattice constant and the band gap of the binary materials are sensitive to the local atomic environments, the descriptor can, up to some extent, capture the effects of atomic arrangements. It additionally includes the very simple features $Z$ and $P$. Also in this case, the components are not highly correlated (maximum $\left| \rho_{i,j} \right|=0.38$).

Also like above, the higher algebraic and informational complexity of $\mathcal{T}_2$ improves the predictive performance. The descriptor depends on band-gap and formation energies, the atomic radius $r_s$, and the dimer distance $d_{2di}$, and these components are almost uncorrelated ($\left| \rho_{1,2} \right|=0.13$). The larger dispersion in the right panels of figure \ref{fig:corr_plots_optimal_descriptors} demonstrates that the learning of $E_\textrm{mix}$ is more difficult than that of $a$, requiring higher descriptor complexity. 

\begin{figure}[h]
\begin{center}
\includegraphics[width=1.0\textwidth]{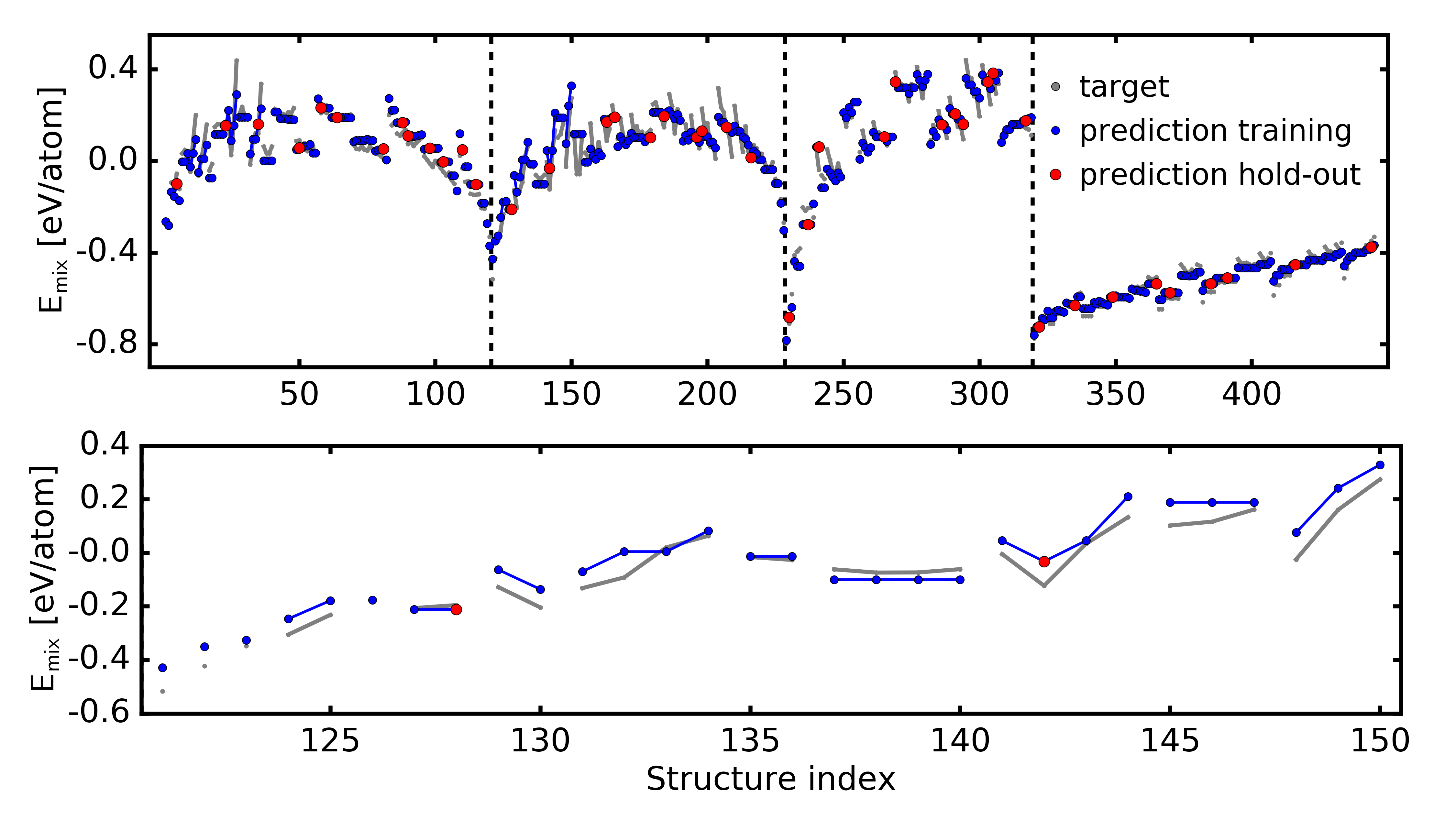}
\caption{Top: Target properties, $E_\textrm{mix}$, and predictions by the optimal 2D descriptor on complexity $\mathcal{T}_2$ for all structures. Samples within a composition are connected by grey lines. Bottom: Same as top panel, but in a narrow range of 30 structures. Here, the capturing of arrangement effects by the model can be seen more clearly.}
\label{fig:optimal_descriptors_data_plot_E0}
\end{center}
\end{figure}

\section{Conclusions} 
\label{sec:conclusions}

We have advanced and applied methods based on compressed sensing that allow one to identify low-dimensional models for predicting materials properties from a potentially huge pool of candidate features. The feature spaces were constructed to contain averages over building blocks of the supercell for being applicable to disordered materials. Moreover, we have derived model selection strategies based on cross-validation to improve the generalizability of the models as compared to approaches that solely use training errors for model selection.

We have evaluated this methodology for predicting the \textit{ab initio} lattice constants and energies of mixing of a dataset of group-IV ternary zincblende compounds.  For extracting a low-dimensional subset of the most relevant features, we have applied LASSO+$\ell_0$ \cite{Ghiringhelli2015,Ghiringhelli2017} and SISSO \cite{Ouyang2018}. Our analysis yields that atomic, bulk, and dimer properties are essential features for this dataset while tetrahedral properties can be discarded. We further found that next-neighbor differences of atomic features and, alternatively, pair features are able to capture arrangement-specific effects of the materials. We have also found a trade-off between larger descriptor dimension and the complexity of the considered features with respect to the algebraic operations. Comparing the feature-selection methods, we confirmed that SISSO reduces correlations between the descriptor components and allows for going beyond the limitations on feature-space size that LASSO+$\ell_0$ suffers from. Although the differences to a simple one-shot strategy \cite{Ghiringhelli2015} are small for the dataset under consideration, our strategy that minimizes the average training error on CV (strategy $\mathcal{S}^\textrm{tr}_\textrm{CV}$) was able to improve descriptor generalizability in terms of hold-out error. We emphasize that the differences between the strategies are small here presumably due to the homogeneity of the dataset. It would thus be desirable to test them in future research on more diverse datasets, for example comprising materials differing in space group or with a larger variety of elements.

\ack
We thank Luca Ghiringhelli for and Matthias Scheffler for fruitful discussions, and Luca Ghiringhelli for valuable feedback to the manuscript. This work has received partial funding from the European Union's Horizon 2020 research and innovation programme through the NOMAD Center of Excellence, grant agreements No. 676580 and 951786.

\bigskip

\bibliographystyle{iopart-num}
\bibliography{main}

\providecommand{\newblock}{}
\begin{thebibliography}{10}
\expandafter\ifx\csname url\endcsname\relax
  \def\url#1{{\tt #1}}\fi
\expandafter\ifx\csname urlprefix\endcsname\relax\def\urlprefix{URL }\fi
\providecommand{\eprint}[2][]{\url{#2}}

\bibitem{Mueller2016}
Mueller T, Kusne A~G and Ramprasad R 2016 {\em Machine Learning in Materials
  Science\/} (John Wiley \& Sons, Inc)

\bibitem{Rupp2012}
Rupp M, Tkatchenko A, M{\"u}ller K~R and Von~Lilienfeld O~A 2012 {\em Physical
  Review Letters\/} {\bf 108} 058301

\bibitem{Behler2011}
Behler J 2011 {\em The Journal of Chemical Physics\/} {\bf 134} 074106

\bibitem{Bartok2013}
Bart\'ok A~P, Kondor R and Cs\'anyi G 2013 {\em Phys. Rev. B\/} {\bf 87} 184115

\bibitem{Seko20141}
Seko A, Takahashi A and Tanaka I 2014 {\em Phys. Rev. B\/} {\bf 90} 024101

\bibitem{Seko2017}
Seko A, Hayashi H, Nakayama K, Takahashi A and Tanaka I 2017 {\em Phys. Rev.
  B\/} {\bf 95} 144110

\bibitem{Sutton2019}
Sutton C, Ghiringhelli L~M, Yamamoto T, Lysogorskiy Y, Blumenthal L,
  Hammerschmidt T, Golebiowski J~R, Liu X, Ziletti A and Scheffler M 2019 {\em
  npj Computational Materials\/} {\bf 5}

\bibitem{Ghiringhelli2015}
Ghiringhelli L~M, Vybiral J, Levchenko S~V, Draxl C and Scheffler M 2015 {\em
  Physical Review Letters\/} {\bf 114} 105503

\bibitem{Ghiringhelli2017}
Ghiringhelli L~M, Vybiral J, Ahmetcik E, Ouyang R, Levchenko S~V, Draxl C and
  Scheffler M 2017 {\em New Journal of Physics\/} {\bf 19} 023017

\bibitem{Butcher2013}
Butcher P~N, March N~H and Tosi M~P 2013 {\em Crystalline semiconducting
  materials and devices\/} (Springer Science \& Business Media)

\bibitem{Ouyang2018}
Ouyang R, Curtarolo S, Ahmetcik E, Scheffler M and Ghiringhelli L~M 2018 {\em
  Physical Review Materials\/} {\bf 2}(8) 083802

\bibitem{Goldsmith2017}
Goldsmith B~R, Boley M, Vreeken J, Scheffler M and Ghiringhelli L~M 2017 {\em
  New Journal of Physics\/} {\bf 19} 013031

\bibitem{Ziletti2018}
Ziletti A, Kumar D, Scheffler M and Ghiringhelli L~M 2018 {\em Nature
  communications\/} {\bf 9} 2775

\bibitem{Musil2018}
Musil F, De S, Yang J, Campbell J~E, Day G~M and Ceriotti M 2018 {\em Chemical
  science\/} {\bf 9} 1289--1300

\bibitem{LeGrain2017}
Legrain F, Carrete J, van Roekeghem A, Curtarolo S and Mingo N 2017 {\em
  Chemistry of Materials\/} {\bf 29} 6220--6227

\bibitem{Ramprasad2017}
Ramprasad R, Batra R, Pilania G, Mannodi-Kanakkithodi A and Kim C 2017 {\em npj
  Computational Materials\/} {\bf 3} 54

\bibitem{Ward2016}
Ward L, Agrawal A, Choudhary A and Wolverton C 2016 {\em npj Computational
  Materials\/} {\bf 2} 16028

\bibitem{Bartok20170}
Bart{\'o}k A~P, De S, Poelking C, Bernstein N, Kermode J~R, Cs{\'a}nyi G and
  Ceriotti M 2017 {\em Science Advances\/} {\bf 3}

\bibitem{Yin1981}
Yin M~T and Cohen M~L 1981 {\em Phys. Rev. B\/} {\bf 24} 6121--6124

\bibitem{Arora2009}
Arora S and Barak B 2009 {\em Computational Complexity: A Modern Approach\/}
  (Cambridge University Press, Cambridge, England)

\bibitem{Tibshirani1996}
Tibshirani R 1996 {\em Journal of the Royal Statistical Society. Series B\/}
  {\bf 58} 267--288

\bibitem{Fan2008}
Fan J and Lv J 2008 {\em Journal of the Royal Statistical Society: Series B\/}
  {\bf 70} 849--911

\bibitem{Ventura2015}
Ventura C~I, Querales~Flores J~D, Fuhr J~D and Barrio R~A 2015 {\em Progress in
  Photovoltaics: Research and Applications\/} {\bf 23} 112--118

\bibitem{Fischer2015}
Fischer I~A, Wendav T, Augel L, Jitpakdeebodin S, Oliveira F, Benedetti A,
  Stefanov S, Chiussi S, Capellini G and Busch K 2015 {\em Optics express\/}
  {\bf 23} 25048--25057

\bibitem{Wendav2016}
Wendav T, Fischer I~A, Montanari M, Zoellner M~H, Klesse W, Capellini G, Von
  Den~Driesch N, Oehme M, Buca D and Busch K 2016 {\em Applied Physics
  Letters\/} {\bf 108} 242104

\bibitem{Hart2008}
Hart G~L and Forcade R~W 2008 {\em Physical Review B\/} {\bf 77} 224115

\bibitem{anote}
For 59 configurations, no stable geometry could be found within the given
  supercell.

\bibitem{Vegard1921}
Vegard L 1921 {\em Zeitschrift f{\"u}r Physik\/} {\bf 5} 17--26

\bibitem{Adachi2009}
Adachi S 2009 {\em Properties of semiconductor alloys: group-IV, III-V and
  II-VI semiconductors\/} vol~28 (John Wiley \& Sons)

\bibitem{Gulans2014}
Gulans A, Kontur S, Meisenbichler C, Nabok D, Pavone P, Rigamonti S, Sagmeister
  S, Werner U and Draxl C 2014 {\em Journal of Physics: Condensed Matter\/}
  {\bf 26} 363202

\bibitem{PerdewWang1992}
Perdew J~P and Wang Y 1992 {\em Physical Review B\/} {\bf 45} 13244

\bibitem{Murnaghan1944}
Murnaghan F 1944 {\em Proceedings of the National Academy of Sciences\/} {\bf
  30} 244--247

\bibitem{Draxl2019}
Draxl C and Scheffler M 2019 {\em JPM\/} {\bf 2} 036001

\bibitem{Draxl2018}
Draxl C and Scheffler M 2018 {\em MRS Bulletin\/} {\bf 43} 676

\bibitem{DentonAshcroft1991}
Denton A~R and Ashcroft N~W 1991 {\em Physical Review A\/} {\bf 43}(6)
  3161--3164

\bibitem{Murphy2010}
Murphy S, Chroneos A, Jiang C, Schwingenschl{\"o}gl U and Grimes R 2010 {\em
  Physical Review B\/} {\bf 82} 073201

\bibitem{Friedmann2001}
Friedman J, Hastie T and Tibshirani R 2001 {\em The elements of statistical
  learning\/} vol~1 (Springer series in statistics New York)

\bibitem{notet2r}
We restrict ourselves to $\mathcal{T}_2^\textrm{\,r}$ due to computational
  limitations in combination with LASSO.

\bibitem{lassol0}
The LASSO preselection step is sub-optimal as compared to solving the exact
  $\ell_0$ problem. This means that, even if the initial feature space is
  enlarged, there is no guarantees that the subspace $\tilde{M}$ left for the
  $\ell_0$ step will lead to a better descriptor.

\end{thebibliography}

\end{document}